\documentclass[twocolumn]{aastex631} %
\usepackage[utf8]{inputenc}
\usepackage{hyperref}
\usepackage{float}
\usepackage{graphicx}
\usepackage{wrapfig}
\usepackage{longtable}
\usepackage{tablefootnote}
\usepackage{threeparttable}
\usepackage{subfigure}
\usepackage{natbib}
\usepackage{hhline}
\usepackage{amsmath}

\begin{document}


\newcommand{\UChicago}{Department of Physics, University of Chicago, 5801 S Ellis Ave, Chicago, IL 60637, USA}
\newcommand{\UCB}{Breakthrough Listen, University of California, Berkeley, 501 Campbell Hall 3411, Berkeley, CA 94720, USA}
\newcommand{\seti}{SETI Institute, 339 Bernardo Ave, Suite 200, Mountain View, CA 94043, USA}
\newcommand{\USQ}{University of Southern Queensland, Toowoomba, QLD 4350, Australia}
\newcommand{\KZA}{University of Malta, Institute of Space Sciences and Astronomy}
\newcommand{\PWJD}{The Breakthrough Initiatives, NASA Research Park, Bld. 18, Moffett Field, CA 94035, USA}
\newcommand{\COL}{Department of Astronomy, Columbia University, 550 West 120th Street, New York, NY 10027, USA}
\newcommand{\Curtin}{International Centre for Radio Astronomy Research, Curtin University, Bentley, WA 6102, Australia}
\newcommand{\GBO}{Green Bank Observatory, 155 Observatory Road, Green Bank, WV 24944, USA}

\newcommand{\tseti}{\texttt{turboSETI}}
\newcommand{\unit}{\dot{\nu}_{\text{unit}}}
\newcommand{\tsnr}{\ensuremath{\rho_{tseti}}}

\correspondingauthor{Carmen Choza}
\email{cgchoza@uchicago.edu \\cgchoza@gmail.com}

\author[0009-0008-0662-1293]{Carmen Choza}
\affiliation{\UCB}
\affiliation{\UChicago}
\affiliation{\seti}

\author[0009-0007-3897-2912]{Daniel Bautista}
\affiliation{\UCB}

\author[0000-0003-4823-129X]{Steve Croft}
\affiliation{\UCB}
\affiliation{\seti}

\author[0000-0003-2828-7720]{Andrew P.\ V.\ Siemion}
\affiliation{\UCB}
\affiliation{\seti}
\affiliation{\KZA}

\author[0000-0002-7461-107X]{Bryan Brzycki}
\affiliation{\UCB}

\author{Krishnakumar Bhattaram}
\affiliation{\UCB}

\author[0000-0002-8071-6011]{Daniel Czech}
\affiliation{\UCB}

\author{Imke de Pater}
\affiliation{\UCB}

\author{Vishal Gajjar}
\affiliation{\UCB}
\affiliation{\seti}

\author{Howard Isaacson}
\affiliation{\UCB}

\author{Kevin Lacker}
\affiliation{\UCB}

\author[0000-0003-1515-4857]{Brian Lacki}
\affiliation{\UCB}

\author[0000-0002-7042-7566]{Matthew Lebofsky}
\affiliation{\UCB}

\author{David H.\ E.\ MacMahon}
\affiliation{\UCB}

\author[0000-0003-2783-1608]{Danny Price}
\affiliation{\UCB}
\affiliation{\Curtin}

\author{Sarah Schoultz}
\affiliation{\UCB}

\author[0000-0001-7057-4999]{Sofia Sheikh}
\affiliation{\UCB}
\affiliation{\seti}

\author{Savin Shynu Varghese}
\affiliation{\seti}

\author{Lawrence Morgan}
\affiliation{\GBO}

\author{Jamie Drew}
\affiliation{\PWJD}

\author{S.\ Pete Worden}
\affiliation{\PWJD}

\title{The Breakthrough Listen Search for Intelligent Life: Technosignature Search of 97 Nearby Galaxies}

\begin{abstract}
The Breakthrough Listen search for intelligent life is, to date, the most extensive technosignature search of nearby celestial objects. We present a radio technosignature search of the centers of 97 nearby galaxies, observed by Breakthrough Listen at the Robert C.\ Byrd Green Bank Telescope. We performed a narrowband Doppler drift search using the \tseti{} pipeline with a minimum signal-to-noise parameter threshold of 10, across a drift rate range of $\pm 4$\,Hz\,s$^{-1}$, with a spectral resolution of 3\,Hz and a time resolution of $\sim 18.25$\,s. We removed radio frequency interference by using an on-source/off-source cadence pattern of six observations and discarding signals with Doppler drift rates of 0. We assess factors affecting the sensitivity of the Breakthrough Listen data reduction and search pipeline using signal injection and recovery techniques and apply new methods for the investigation of the RFI environment. We present results in four frequency bands covering 1 -- 11\,GHz, and place constraints on the presence of transmitters with equivalent isotropic radiated power on the order of $10^{26}$\,W, corresponding to the theoretical power consumption of Kardashev Type II civilizations.

\end{abstract}


\section{Introduction}
The Search for Extraterrestrial Intelligence (SETI) seeks to answer one of the most persistent and profound questions faced by modern science: does life exist elsewhere in the universe? The earliest searches for evidence of technologically advanced life, or ``technosignatures'', began in the 1950s and 60s with the work of \citet{cocconi:1959} and \citet{project_ozma} and were restricted to the domain of radio communications over narrow bandwidths. Techniques have since expanded to allow for varied and innovative approaches to the search for extraterrestrial life, including in-situ sampling by spacecraft \citep{Dragonfly_concept, biosignatures_2018}, the remote investigation of exoplanet atmospheres for signatures produced by biological phenomena \citep{Thompson__2022}, and technosignature surveys at optical \citep{Zuckerman_2023} and infrared wavelengths. Radio technosignature searches have remained popular due to the low extinction of radio waves in the interstellar medium, which allows a vast range of distances and objects to be observed. These searches have advanced to cover far wider bandwidths and larger numbers of targets than ever before \citep{2020arXiv200611304L, 2021PASP..133f4502C} and continue to be powerful tools for SETI. Successful detection of artificially structured radio emission would unambiguously imply the existence of extraterrestrial, intelligent life, due to the lack of natural, astrophysical confounders for these signal morphologies.

\begin{figure*}
    \centering
    \includegraphics[width=2\columnwidth]{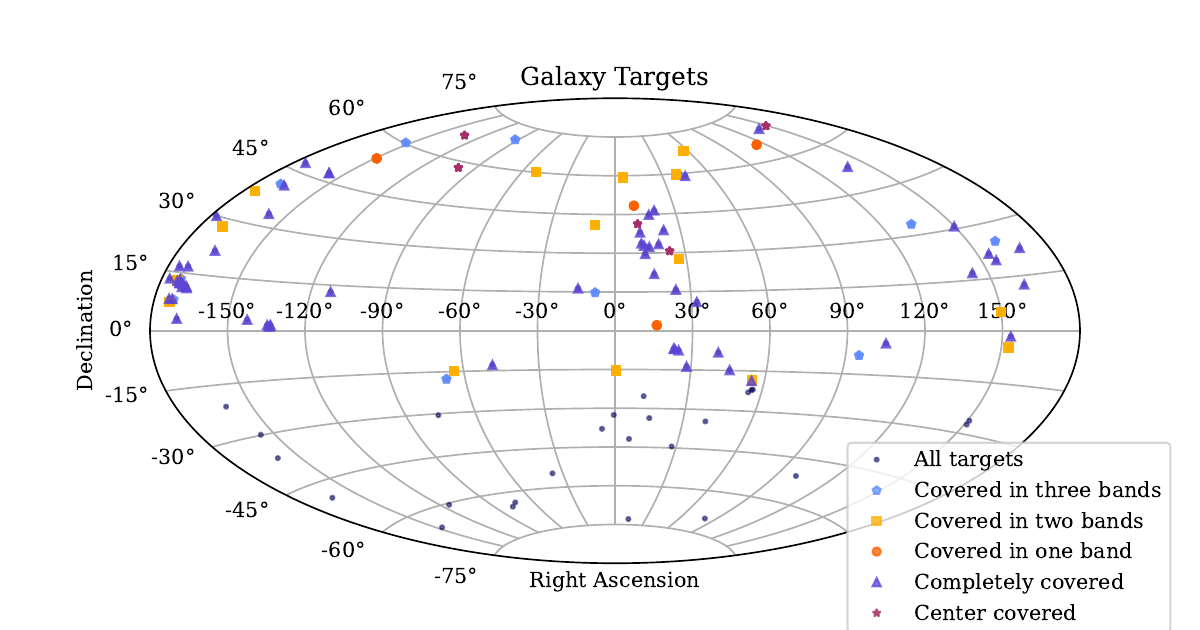}
    \caption{Sky map of galaxy targets. The colored squares indicate whether the full source is covered by the GBT beam at all four bands (see Table \ref{tab:receivers}), three bands, two, one, or none. The blue dots are galaxy targets in the southern hemisphere that are part of the Parkes Observatory sample for observation by Murriyang and are not discussed further here. At least one cadence was taken per source per band (Appendix \ref{app:targets}). All ``on'' pointings in a given cadence were aimed at the center of the target galaxy.}
    \label{fig:sky_map}
\end{figure*}

Since 2015, the Breakthrough Listen (BL) Initiative has constituted humanity's most significant effort to date to search our skies for extraterrestrial technosignatures \citep{Worden:2017}. BL operates at a wide range of wavelengths spanning optical to radio, using a wide variety of telescopes including the 100 m Robert C.\ Byrd Green Bank Telescope (GBT) in West Virginia, the Automated Planet Finder (APF) in California, and the CSIRO Parkes `Murriyang' 64-m radio telescope in Australia. To date, BL has conducted numerous technosignature searches of nearby stars \citep{Price:2020}, known exoplanets \citep{Franz_2022,Traas}, and the Galactic Plane \citep{Gajjar-2021}, in addition to a variety of other targets. 

The great distances to the nearest galaxies make successful detection of emission much more difficult than between nearby stars, requiring signals on the order of $10^{10}$ more powerful to be sent in order for us to detect them on Earth. However, galaxies provide the opportunity to observe billions of stars in a single observation and to sample the brighter end of the artificial emitter luminosity function. On the Kardashev scale \citep{Kardashev:1964}, these ten orders of magnitude correspond to the leap between a Kardashev I and Kardashev II civilization for typical sensitivities in our GBT observations. A Kardashev Type I civilization would be able to utilize the full energy resources of its host planet, whereas a Kardashev Type II civilization would be able to utilize the full energy of its host star. At present, humankind has yet to reach a ranking of Kardashev Type I \citep{Zhang_Yang_Luo_Fan_2023}.

A successful detection of an artificial signal coming from another galaxy could not only prove the existence of an extraterrestrial intelligence but also the existence of a civilization that possesses technology far greater than that of humankind. Such a signal may be vanishingly rare. Despite the immense power required to produce a signal that we could observe with our current sensors, it is feasible that technology could advance to sufficient levels, and the opportunity to observe billions of objects with varying size, construction, and age makes such a search valuable and complements searches of nearer objects.

Previous SETI studies have turned to maximizing stellar number to expand the search parameter space, but few have conducted searches for narrowband radio signals in galaxies with high resolution. Even targeted searches of nearby stars contain significant ``bycatch'' of more distant stars; \citet{Wlodarczyk-Sroka:2020} extended earlier BL searches \citep{Enriquez,Price:2020} significantly by calculating the stellar bycatch. Surveys within the Milky Way can be specifically designed to capture large numbers of stars by targeting regions of high stellar density \citep{Gajjar-2021, Tremblay:2020}, and \citet{GrayMooley:2017dr} observed the galaxies M31 and M33 at 21cm using the Jansky VLA with $<1$\,kHz spectral resolution, sampling $\sim 10^{12}$\, stars and establishing one of the deepest constraints on the luminosity function of extraterrestrial transmitters in the Local Group. \citet{GarrettSiemion:2022} and \citet{Uno_2023} applied a similar analysis to identify serendipitous observations of extragalactic objects in GBT data. Their analyses of galaxies identified in those fields sampled $\sim 10^{11}$\, and over $10^{13}$ stars, respectively, with a spectral resolution of $3$\,Hz, though the greater average distance of their galaxy targets leads to correspondingly higher power requirements and lower sensitivity than the search of nearby galaxies that we present here.

\subsection{The Green Bank Telescope}

All data analyzed in this work were collected using the GBT, a 100 meter telescope located in West Virginia, USA and operated by the Green Bank Observatory. The telescope is capable of observing at frequencies between 0.2 --- 116\,GHz, depending on the receiver selected. We collected data for this work using four receivers spanning the range from 1 --- 11\,GHz: the L-band receiver from 1.10 --- 1.90\,GHz; the S-band receiver from 1.80 --- 2.70\,GHz; the C-band receiver from 4.00 --- 7.80\,GHz; and the X-band receiver from 7.80 --- 11.20\,GHz. A user-selectable notch filter blocks the L-band between 1.20 and 1.34\,GHz, which is used by BL observations. A permanent notch filter blocks the S-band between 2.3 and 2.36\,GHz. Table \ref{tab:receivers} provides band information as well as the $\text{EIRP}_{min}$ per receiver, calculated with each receiver's system temperature and a signal-to-noise (S/N) threshold of 33 (see Section \ref{turboseti_performance}).

\subsection{Galaxy sample} \label{galaxy_sample}

 The BL nearby galaxy sample, detailed by \cite{isaacson:2017}, consists of 123 targets. Our sample of 97 is made up of those targets from \citeauthor{isaacson:2017} with declinations above $-20 \degr$; targets below this declination are observed by the BL program at Parkes. Together, they compose a morphological-type-complete set of the nearest ellipticals, dwarf spheroidals, irregulars, and spiral galaxies like our own. The sky positions of the targets are shown in Figure~\ref{fig:sky_map}.

\begin{figure}
    \centering
    \includegraphics[width=\columnwidth]{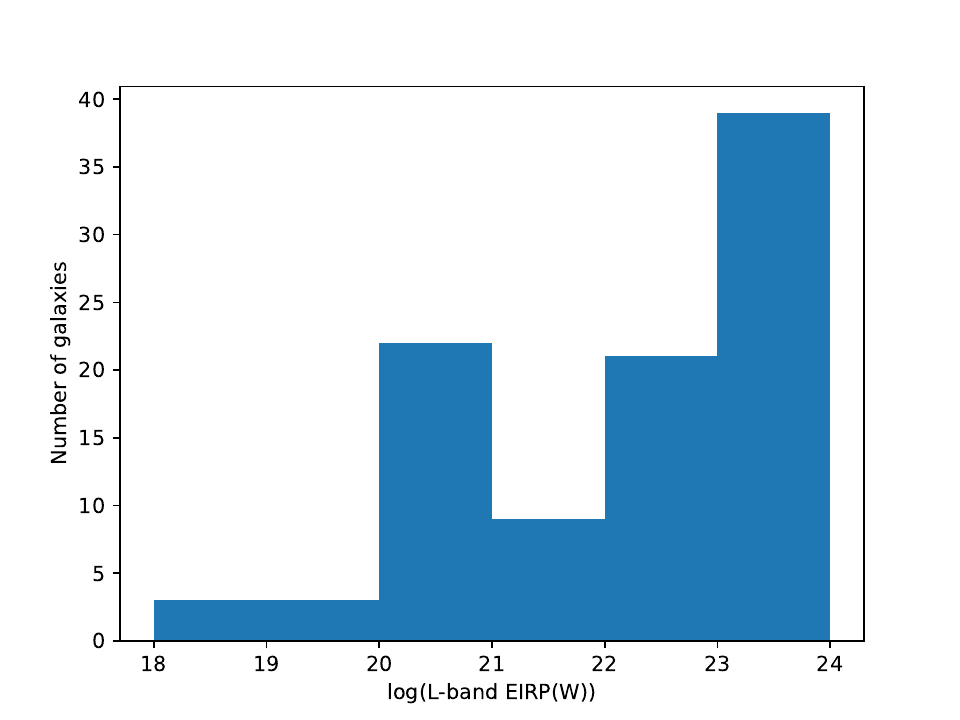}
    \caption{Histogram of the 97 galaxies in our sample binned by the equivalent isotropic radiated power required for a transmitter (with a bandwidth narrower than 2.79\,Hz) in a given galaxy to be detectable in 300\,s in our GBT observations. The powers are computed using the characteristics of the GBT's L-band receiver. The S, C, and X band receivers have slightly worse sensitivity and hence slightly higher EIRP values.}
    \label{fig:lband_eirp}
\end{figure}

Distances to the 97 galaxies in the GBT sample range from 60\,kpc (the Ursa Minor Dwarf galaxy) to 29.2\,Mpc (NGC\,5813). Therefore, even though each target is observed with approximately uniform sensitivity, the minimum equivalent isotropic radiated power (EIRP) for a transmitter in NGC\,5813 to be detectable by our observations is five orders of magnitude brighter than for a transmitter in the Ursa Minor Dwarf. A histogram of minimum EIRPs for the 97 galaxies in our sample is shown in Figure~\ref{fig:lband_eirp}.

\begin{figure}
    \centering
    \includegraphics[width=\columnwidth]{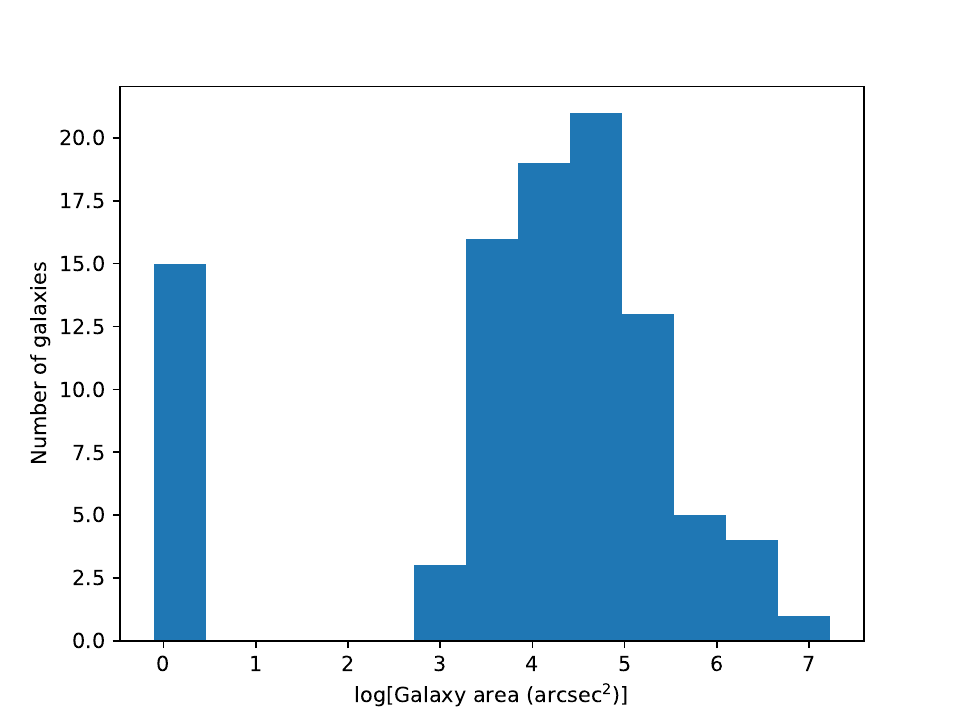}
    \caption{Sky area of the 97 galaxies in our sample, as determined from the major and minor axis values cataloged in the NASA Extragalactic Database.}
    \label{fig:galsize}
\end{figure}

\begin{figure*}
     \centering
     \includegraphics[width=0.5\linewidth]{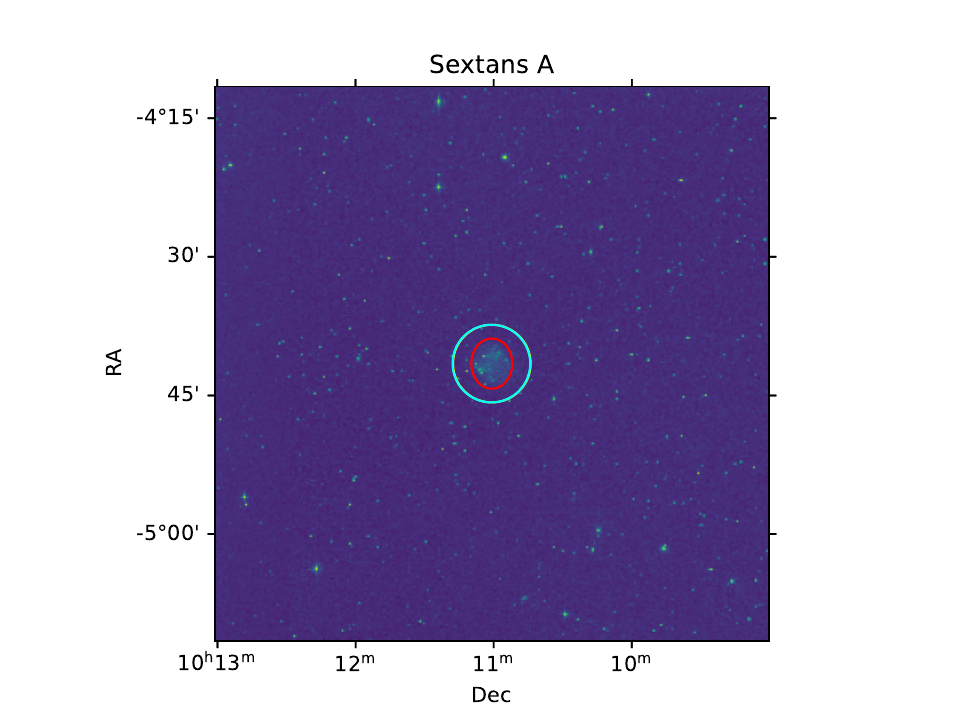}%
     \includegraphics[width=0.5\linewidth]{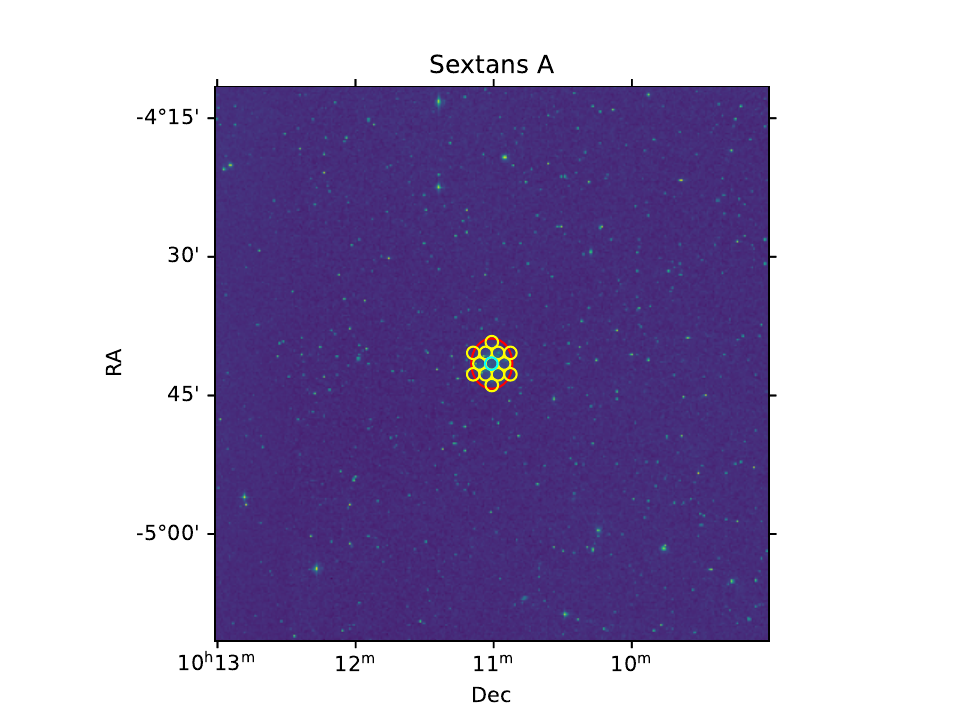}
     \includegraphics[width=0.5\linewidth]{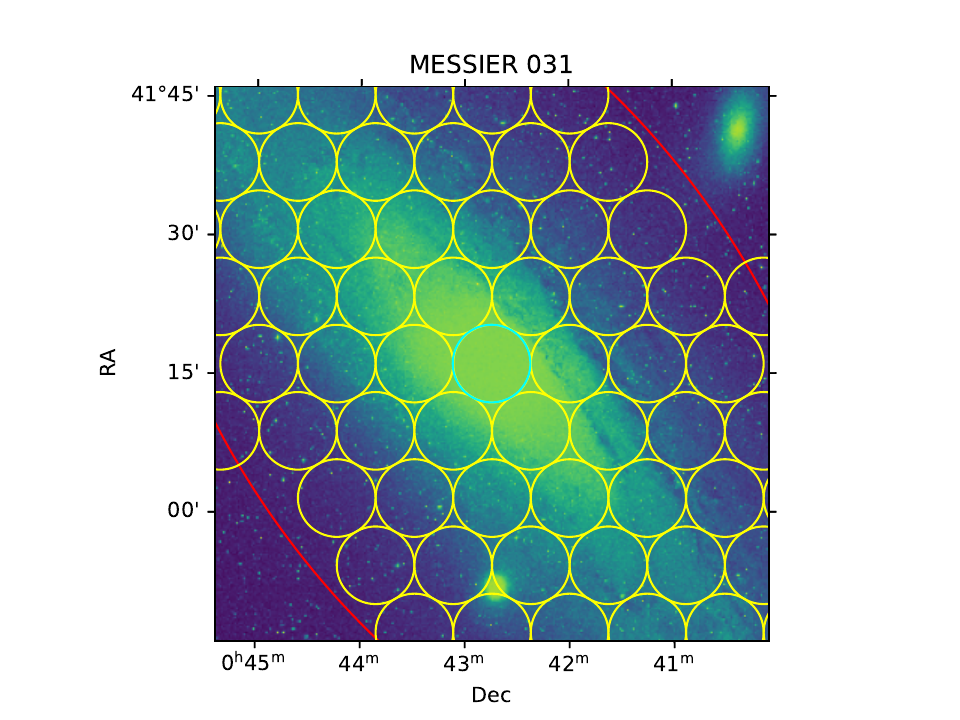}%
     \includegraphics[width=0.5\linewidth]{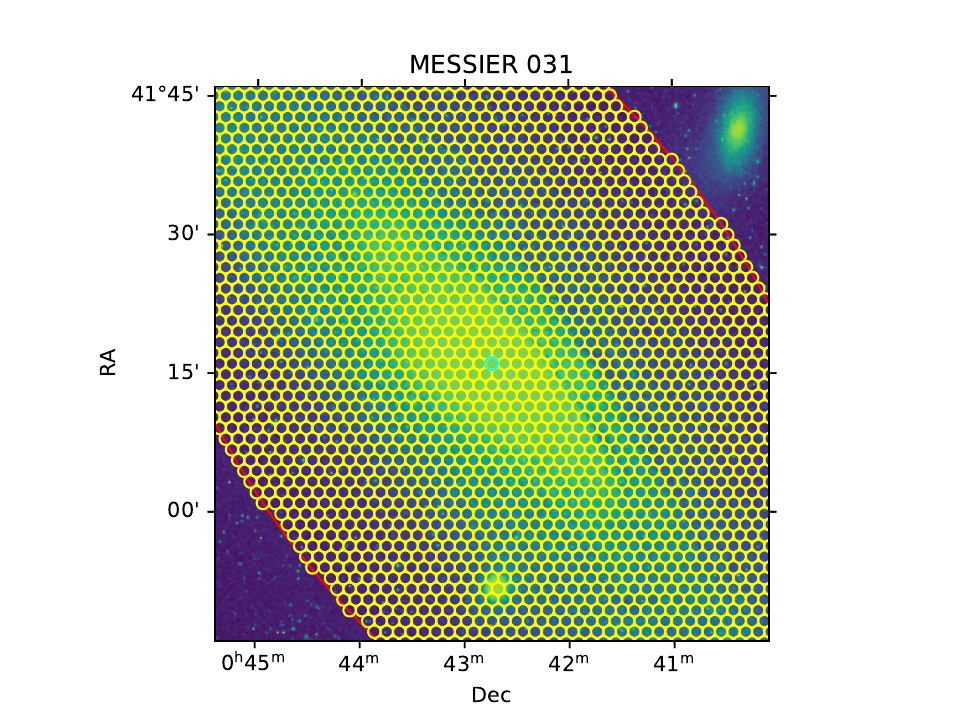}
     \caption{Examples of pointings needed to tile extended galaxies at L-band (left column) and at X-band (right column); S-band and C-band are intermediate in beam size between these two extremes. Small galaxies such as Sextans\,A (top row) can be covered by a single pointing at some bands but may require multiple pointings to cover at other bands, whereas extended galaxies (including M\,31, the largest galaxy in our sample) require multiple pointings at all bands. Yellow circles show the pointings, tiled so that they touch at the half-power point for a given receiver, required to fill the optical extent of the galaxy (red ellipse, determined from the major and minor axis and position angle from the NASA Extragalactic Database). The sample analyzed here consists only of the central pointing for each galaxy, shown here in cyan.}
     \label{fig:pointings}
\end{figure*}

The half-power beamwidth of the GBT receivers is a function of frequency; the L-band receiver covers a sky area almost 40 times larger than the X-band receiver. Of the 97 galaxies in our sample, 90 are small enough in angular extent that they are entirely covered within a single L-band pointing (see Figures~\ref{fig:sky_map} and \ref{fig:galsize}). For the remaining galaxies, we observe a single central pointing, and some of the associated stars fall outside of the half-power beam width. In the future we plan to tile the outer regions of some of these galaxies with further pointings (see Figure~\ref{fig:pointings}). A total of 6161 pointings are needed to completely cover all of the galaxies at X-band, which includes 3029 pointings dedicated to M\,31 alone (the most extended galaxy in our sample), making it impractical to cover in a reasonable amount of time using our standard observation procedure on the GBT.

\begin{center}
\begin{table*}
\begin{threeparttable}
    \caption{Survey Parameters}
    \begin{tabular}{c c c c c c c c}
        \hline
        \hline
        Receiver & Frequency & Cadences & Hits & Events & CWTFM\tnote{a} & EIRP$_{\textrm{min}}$  & Transmitter \\
        & [GHz] & & & & & [10$^{24}$ W] \tnote{b} &  Limit [\%]\tnote{c} \\
        \hline
        L & 1.10 - 1.90 & 100 & 2186151 & 288 & 3793 & 2.31 & 3.7\\
        S & 1.80 - 2.70 & 156 & 853434 & 250 & 2828 & 2.61 & 3.7\\
        C & 4.00 - 7.80 & 106 & 1418492 & 650 & 3059 & 3.1 & 3.7\\
        X & 7.80 - 11.20 & 97 & 1576236 & 331 & 3686 & 3.46 & 3.7\\
        \hline
        Total & 1.10 - 11.20 & 459 & 6034313 & 1519 & - & - & - \\
        \hline
    \end{tabular}
    \begin{tablenotes}
        \footnotesize
        \item[a] Continuous Waveform Transmitter Figure of Merit (CWTFM) is a figure of merit that describes the likelihood to find a signal above the EIRP$_{\textrm{min}}$ for that receiver.
        \item[b] Minimum Equivalent Isotropic Radiated Power (EIRP$_{\textrm{min}}$) is a measure of the minimum necessary omnidirectional power of a transmitter at each receiver to be detected. 
        \item[c] The transmitter limit is the maximum percentage of galaxies in each frequency range that \textbf{possess} a transmitter, given that we find no signals for 97 targets (trials). This figure is discussed further in Section \ref{Doppler Search}.
    \end{tablenotes}
    \label{tab:receivers}
\end{threeparttable}
\end{table*}
\end{center}

\subsection{Kardashev Type II Civilizations}

The great distances to other galaxies pose a formidable challenge for intergalactic communication. For an isotropic beacon to be detectable from another galaxy, it would need to be transmitted with power equivalent to the output of an entire star. Even a Kardashev Type II civilization, with technology capable of capturing sufficient power, would likely still face the physical limits of conventional dish antennas. Waste heat is an inevitable byproduct of any substantial broadcasting antenna, necessitating efficient heat dissipation mechanisms to prevent structural failure. The amount of waste heat that an ideal radiator of area A and temperature T can dissipate can be calculated using the Stefan-Boltzmann law:

\begin{equation}
\label{eq:waste_heat}
L_{\text{waste}} = \sigma_{\text{SB}} \cdot T^4 \cdot A
\end{equation}

\noindent For an order-of-magnitude estimate, if we assume that about half the power is dissipated within the structure as waste heat \citep{bonsall_gbt_2019}, an antenna capable of transmitting 1 $L_{\sun}$ would require a radiator capable of dissipating an equivalent quantity of waste heat:

\begin{equation}
L_{\text{waste}} = 74.5 \, \text{TW} (\frac{T}{6,000 \, \text{K}})^4 (\frac{A}{1 \, \text{km}^2})\\
\label{eq:norm_waste_heat}
\end{equation}

No currently known solid material can withstand temperatures of 6,000\,K, so such a radiator would have to be sufficiently large to radiate 1 $L_{\sun}$ without itself melting. Since present-day Earth electronics generally function at Earth temperatures or cooler, we can assume a radiator would have to cool the antenna to temperatures around 300K. The surface area of an ideal radiator capable of emitting an equivalent power to the sun at a temperature of 300K would then be:

\begin{equation}
A_{\text{radiator}} = \frac{L_{\text{waste}}}{74.5\,\text{TW}(\frac{300}{6000})^{4}} = 8.23 \times 10^{17}\,\text{km}^2,
\end{equation}

\noindent similar in size to the surface area of the sun. While such a construction would require colossal expense and effort, a Kardashev Type II civilization would require a device of comparable or greater size to collect stellar power. For a sun-like star and Earth-like electronics, a Dyson swarm would have to be approximately the size of a sphere of radius 1 AU, and a civilization capable of constructing such a system would therefore likely be able to replicate the feat. However, any construction of significant scale could face gravitational instability.

 A more feasible approach involves the construction of a vast array consisting of numerous smaller transmitters operating in concert. The accumulation of coordinated, phase-aligned, weaker signals from many less powerful beacons could also produce a sufficiently bright radio signal. In the case of a system powered by a sun-like star, the collectors themselves could serve the triple purpose of energy collection, transmission, and heat dissipation. A beacon would then take the form of a complex of transmitters, with the scale of the array varying from planetary dimensions, if highly focused, to encompassing an entire solar system, if isotropically radiating. Arrays of highly beamed laser or microwave emitters have been proposed as means of interstellar propulsion \citep{Parkin_2018}, and some propulsion signatures may be detectable from interstellar distances \citep{Lingam_2017}.

We have based our considerations on isotropically emitting, continuous beacons, but a beamed transmitter would require far less power to achieve an equivalent EIRP. By tightly focusing the beam, the Effective Isotropic Radiated Power (EIRP) of a given antenna can be significantly increased, harnessing the high gain of large dish antennas. A more targeted array reduces the power requirements considerably. In order to encompass the region within 10\,kpc of the Galactic Center from a distance of 10\,Mpc, the required beam coverage ($\Omega$) would amount to approximately $3.14\times10^{-6}\,\text{sr}$ or 37 square arcminutes. Should the transmitter solely target that portion of the sky, the Equivalent Isotropic Radiated Power (EIRP) would be 4 million times brighter than its actual luminosity. Then the actual trasmitting power of the array is approximately $2.5\times10^{-7}$ times that of an isotropic beacon, or $1.9\times10^{20}$\,W. The corresponding radiator (again applying Eq. \ref{eq:norm_waste_heat}) necessary to keep the transmitter system operating would need to have a collective surface area of $4.11\times10^{11}\,km^{2}$, several times the surface area of Jupiter but much smaller than the Sun. Still, even a system designed to illuminate the entire Milky Way, whether simultaneously or intermittently, would be much more affordable to construct, and would represent a significantly smaller fraction of a civilization's power budget. \citet{GrayMooley:2017dr} suggest that the Milky Way's status as the second largest in the Local Group could make it a prime target for ETI attempting to communicate, and provide a detailed treatment of some lower-power possibilities.

\section{Observations}

The BL backend on the GBT \citep{macmahon:18, lebofsky:19} enables the storage and analysis of greater volumes of SETI data than ever before possible. BL observations at the GBT employ a ``cadence'' strategy, whereby a target in the sample (an ``on'' source) is observed for five minutes, and then an offset location is observed several beamwidths from the target (an ``off" source). The on/off pattern is repeated three times with three separate ``off'' pointings observed for 5 minutes each, resulting in a 30-minute \mbox{ABACAD} cadence \citep{lebofsky:19}.  Comparing the ``on'' and ``off'' scans allows us to discriminate between signals of interest and close-by radio frequency interference (RFI), as signals that are not localized on the sky are likely to be accidental detections of anthropogenic sources. Signals of interest are required to be present in all of the three ``on'' scans and none of the ``off'' scans to ensure that the signal is both localized and continuous. Signals not accelerating relative to the Earth, i.e. those with Doppler drift rates of 0\,Hz\,s$^{-1}$, are rejected as they most likely correspond to terrestrial radio interference \citep{Sheikh-2019}.

The 459 cadences analyzed in this work represent 139.7\,TB of fine-frequency resolution spectrograms stored in HDF5 format, recorded by the BL backend at the GBT. The number of cadences is greater than one per object per band since some objects were observed multiple times over the course of data collection. Only cadences that had complete data across the frequency ranges in Table \ref{tab:receivers} were selected; a summary of the data sample is provided in Appendix \ref{app:targets}.
 
 The BL backend records spectral data in 187.5\,MHz frequency chunks, with each chunk downsampled and channelized into multiple time and frequency resolutions by a separate compute node. The fine-frequency resolution spectra have a frequency resolution of $\sim 2.79$\,Hz and a time resolution of $\sim 18.25$\,s. The data products are described in detail by \citet{lebofsky:19} and \citet{macmahon:18}. Before early 2021, spectra from separate compute nodes were spliced together to form broadband spectra. Since early 2021, files were instead left in their unspliced form to facilitate transfer and parallel processing. The splicing state does not have any effect on the following analysis. The sample spans five years from early 2018 to late 2022, totalling 229\,h of observation time, and includes both spliced and unspliced data. Because of a mechanical issue with the GBT in early 2023, we were unable to obtain complete cadences for six objects at X-band; for these targets, we use X-band observations for which one compute node failed to record data during the observation in at least one scan, leading to a gap of 187.5\,MHz in the spectrum.

\section{Doppler Search} \label{Doppler Search}

Each cadence was analyzed using the BL \tseti{} pipeline \citep{turboSETI} to search for linearly-chirped narrowband signals. The \tseti{} method \texttt{FindDoppler} identifies narrow-band Doppler-drifting signals in the filterbank files. The bulk of the sample was processed using \texttt{seticore}\footnote{https://github.com/lacker/seticore}, a high-performance implementation of core \tseti{} algorithms; the two pipelines produce the same outputs. The \tseti{} algorithm requires a minimum signal-to-noise (S/N) ratio and maximum drift rate range to limit computational expense and filter noise.
Doppler shift has units of Hz, and we describe Doppler drift rates in units of Hz\,s$^{-1}$. However, to generalize across the search bandwidth, recent SETI studies such as \citet{Sheikh-2019} have taken to normalizing the drift rate by the ``rest frequency'', conventionally the center frequency or upper frequency of the observing bandwidth, in order to discuss a normalized drift rate that is independent of transmission frequency. A normalized drift rate will correspond to the spectrum of ``true'' drift rates dependent on the transmission frequency that would be produced by an emitter traveling at a single relative acceleration. \citet{Sheikh-2019} recommend that drift rates in the range of $\pm 200$\,nHz, equivalent to a drift rate of $\pm 200$\,Hz\,s$^{-1}$ at a center frequency of 1\,GHz, should ideally be used to search for narrowband signals in order to account for all possible drift rates produced by observed objects and exoplanets. Past studies have acknowledged this recommendation but chosen more conservative drift rates to target Earth-like planets or limit computational requirements. A recent study by \citet{Li_2023} extended the work done by \citet{Sheikh-2019} to a larger sample and modeled the drift rate distribution of exoplanets using NASA Exoplanet Archive (NEA) data as well as a simulated population to account for ease-of-detection bias. The authors find that a maximum drift rate of $\pm 47$\,nHz would be sufficient to catch 99\% of drift rates produced by the accelerations of known NEA exoplanets, while the simulated population reduces maximum drift rate requirements at the 99 percent level to the $\pm 0.5$\,nHz range. At our lowest observing frequency of 1.1\,GHz, this value corresponds to a non-normalized drift rate of 0.55\,Hz\,s$^{-1}$; at the highest observing frequency we cover, the non-normalized drift rate is 5.6\,Hz\,s$^{-1}$. These results suggest that our chosen drift rate range covers the majority of the parameter space of interest.

Following \citet{Price:2020} and \citet{Traas}, we adopted a minimum\footnote{\tseti{}'s dechirping efficiency is lower for high drift rate signals, resulting in a higher effective S/N limit. See \citet{Gajjar-2021}, and Section 4 of this work.} signal-to-noise threshold parameter (\tsnr{}) of 10; however, we note that this is different from the signal-to-noise threshold of our search and will be discussed further in Section \ref{turboseti_performance}. We chose a drift rate range of $\pm 4$\,Hz\,s$^{-1}$, again for consistency with past BL searches. This drift rate covers the range recommended by the exoplanet population simulations of \citet{Li_2023} at L, S, and C bands, and 70\% of that range at X band. We refer to any signal detected by the \texttt{FindDoppler} or \texttt{seticore} algorithm in a single observational spectrum as a ``hit", and do not record zero-drift hits. We pass the recorded hits through the \texttt{find\_event} method of \tseti{}, which removes any hits for which an ``off" observation contains a hit in the range

\begin{equation}
\label{eq:discard_range}
\nu_{off} = \nu \pm |\dot \nu| \times 2(\Delta \textrm{T})
\end{equation}

(where $\nu$ is the central frequency, $\dot \nu$ is the drift rate, and $\Delta \textrm{T}$ is the total time of an observation) and compares remaining hits across the ``on" scans in a cadence. Any collection of three hits forming a signal which appears in all of the ``on" scans and none of the ``off" scans we flag as an ``event". Events were then plotted with \tseti{}'s \texttt{plot\_event} method and inspected visually for RFI and false positives remaining after the \texttt{find\_event} step. 

\subsection{turboSETI Performance} \label{turboseti_performance}

As in other BL searches, we focus on continuous, narrowband Doppler-drifting beacon signals. The \tseti{} pipeline searches for linearly chirped signals in time-frequency spectra using an incoherent Taylor ``tree deDoppler" algorithm as described by \citet{Siemion:2013}. The tree summation method is computationally advantageous and allows fast analysis of large data volumes, but, as it is an incoherent method, the detection efficiency for complex or high-drift signals is influenced by the time-frequency resolution of the data products. Signals with drift rates higher than the ``unit drift rate", defined as $\unit{} = \frac{\Delta f}{\Delta t}$, where $\Delta f$ is the fine-frequency channel bandwidth and $\Delta t$ is the length of a single time integration, will smear across multiple frequency channels in the same time integration, resulting in a reduced intensity in the Doppler-corrected integrated spectrum. For the BL fine-frequency resolution data products, $\unit \sim 0.153$\,Hz\,s$^{-1}$. 

Recent work has detailed resolution-dependent limitations of the \tseti{} algorithm. \citet{Margot-2021} conducted a signal injection and recovery analysis quantifying \tseti{} performance in noise-free constant-power synthetic spectra, confirming that \tseti{} reaches peak efficiency at drift rates $\dot{\nu} \leq \unit$, and decreases as $1/N$ with increasing drift rate, where $N$ is the number of frequency channels a drifting signal spreads in $\Delta t$. Modifications to our search algorithms to compensate for these effects are ongoing. To inform these efforts, we conducted signal injection and recovery tests to investigate \tseti{}'s performance in real data as a function of signal intensity and drift rate and its response to detailed simulated technosignatures.

\subsubsection{Signal Generation and Injection} \label{subsec_signal_generation}

\begin{figure}
    \centering
    \includegraphics[width=\columnwidth]{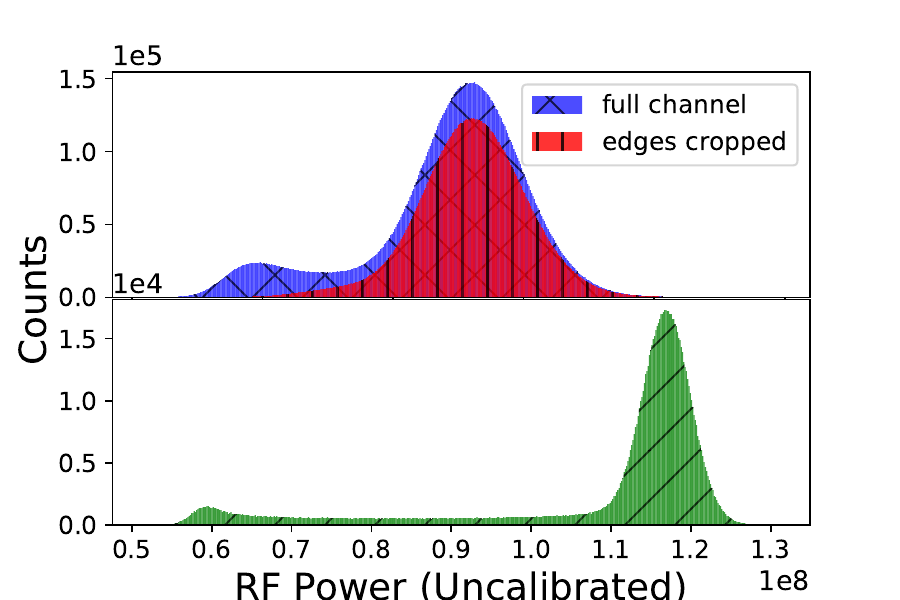}
    \caption{Top: The data distribution of a low-RFI coarse channel, with DC (direct current) spike removed, overlaid by the same distribution with the coarse channel roll off cropped out. Bottom: the same data distribution time-summed, as is done in \tseti{} before noise calculation.}
    \label{fig:coarse_channel_noise}
\end{figure}

Signals were simulated using \texttt{setigen} \citep{Brzycki_2022}, a package developed for the generation of synthetic radio observations in time-frequency spectra and time-series voltage data. \texttt{setigen} allows sophisticated control of signal morphology, grounded in physical principles. For this analysis, we make use of the spectrogram module method \texttt{add\_signal} to inject linearly drifting, time-constant, narrow-band signals. Signals were injected with a $\text{sinc}(f)^2$ frequency profile to simulate the effects of our pipeline's Fast Fourier Transform (FFT) on a sinusoidal signal in voltage data. For ease of processing, we split base files into $N_{coarse}$ $\sim 2.93$\,MHz wide frames according to coarse channelization in the data products, ran \tseti{} on the frame with S/N = 10 and a drift rate range of $\pm 4$\,Hz\,s$^{-1}$, inserted $N_{in}$ signals, and re-ran \tseti{} on the saved frame containing synthetic data.

\subsubsection{Coarse Channel Response and Noise Calculations}\label{noise_calculations}

We first measured the pipeline's efficiency at detecting signals at or near the S/N value, \tsnr, specified by the user when running \tseti{}, which revealed discrepancies between injected and recovered signal to noise ratios. The disagreement on S/N value between \tseti{} and \texttt{setigen} arises due to the different methods of noise calculation utilized by the two software packages, as well as bias in injection and noise calculation stages due to the coarse channel response. 

\begin{figure}
    \centering
    \includegraphics[width=\columnwidth]{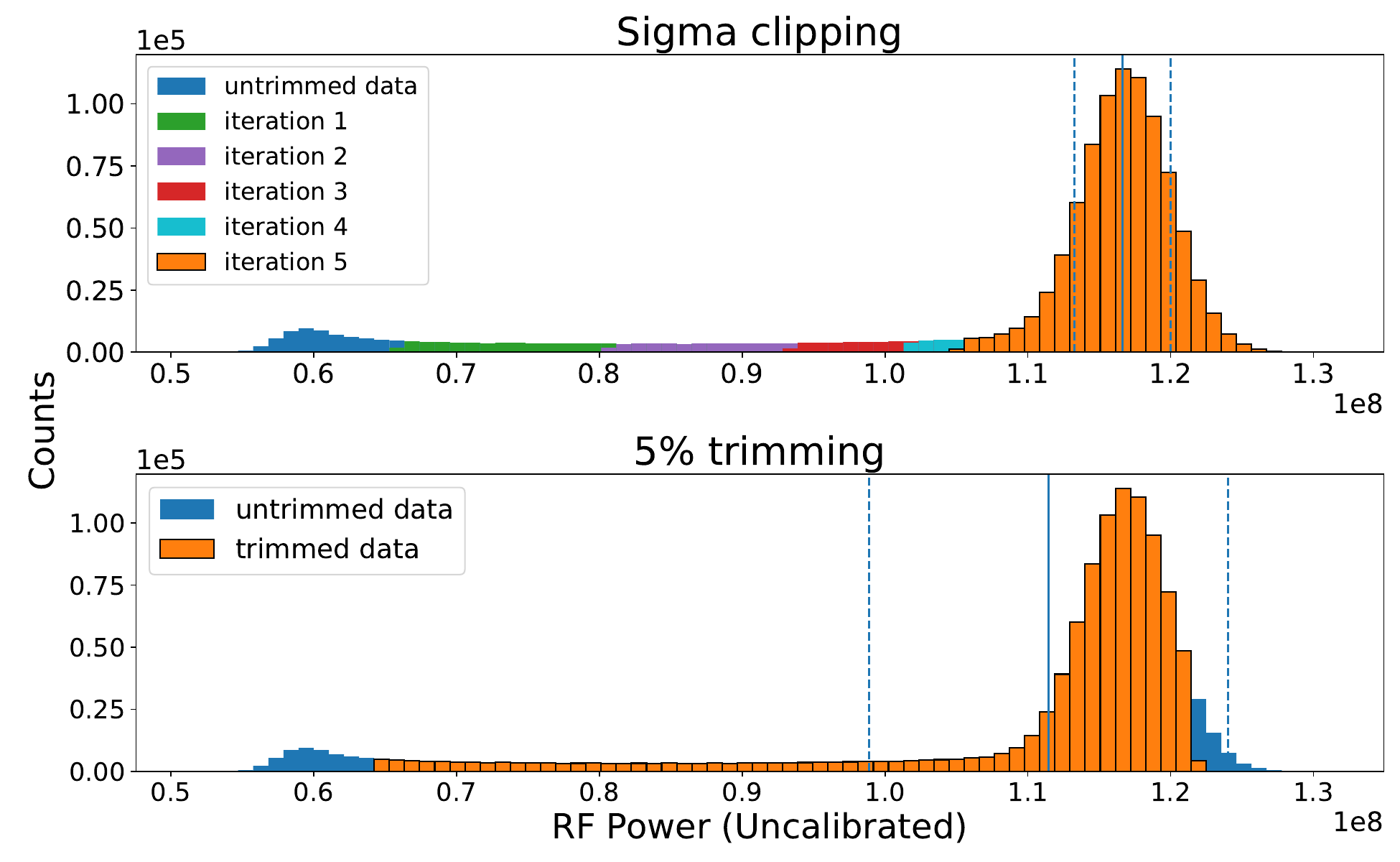}
    \caption{Top: The resulting data distribution of a coarse channel treated with a $3\sigma$ threshold, five-iteration sigma clipping process. Histograms for the data remaining at each iteration are laid over the initial, untrimmed data. Bottom: A histogram of the data remaining after trimming the brightest and dimmest 5\% of pixels over the histogram of the untrimmed coarse channel. The vertical solid and dotted lines mark the mean and $\pm 1\sigma$ from the mean of the final trimmed distribution. The sigma clipping method trims the ``tail'', while the 5\% method crops out only a small portion of the tail as well as a portion of samples within $1\sigma$ above the mean.}
    \label{fig:clipping_methods}
\end{figure}

The roll-off in sensitivity due to the coarse channelization process is well-known, but its impact on \tseti{} noise calculation has not been fully appreciated. Figure \ref{fig:coarse_channel_noise} shows a histogram of counts across a coarse channel, which displays a bimodal profile. One peak corresponds to the data distribution in the channel plateau, while the lower intensity peak corresponds to the data distribution of the coarse channel roll off. After summing the spectrogram along the time axis with no dedrifting (integrating across a full observation), the overlap of the two distributions creates a long low-intensity ``tail'', while the Gaussian representing the data distribution in the channel plateau narrows. \tseti{} version 2.3.2 searches a coarse channel at a time and calculates the S/N of a signal by time-summing the data array, discarding the dimmest and brightest 5\% of pixels, then calculating the median and standard deviation directly from the remaining distribution using \texttt{numpy} \citep{harris2020array} methods. In contrast, \texttt{setigen} applies a sigma clipping function which iterates five times at a $3\sigma$ level to remove outliers and arrive at a single-mode noise mean. Figure \ref{fig:clipping_methods} shows the application of each method to one low-RFI coarse channel, with trimmed data overlaid on the untrimmed histogram. The 5\% method discards only a small fraction of the tail, as well as a portion of samples within $1 \sigma$ of the mean of the remaining data, whereas the sigma clipping approach eliminates most if not all of the tail and returns a mean centered on the higher noise peak and a significantly smaller standard deviation. The factor of difference in the standard deviations produced by the two methods varies according to the noise and RFI present in a given frame, but for 1280 coarse channels across a full X-band observing bandwidth from 7500 MHz to 11250 MHz, has a mode of 3.25.

\begin{figure}
    \centering
    \includegraphics[width=\columnwidth]{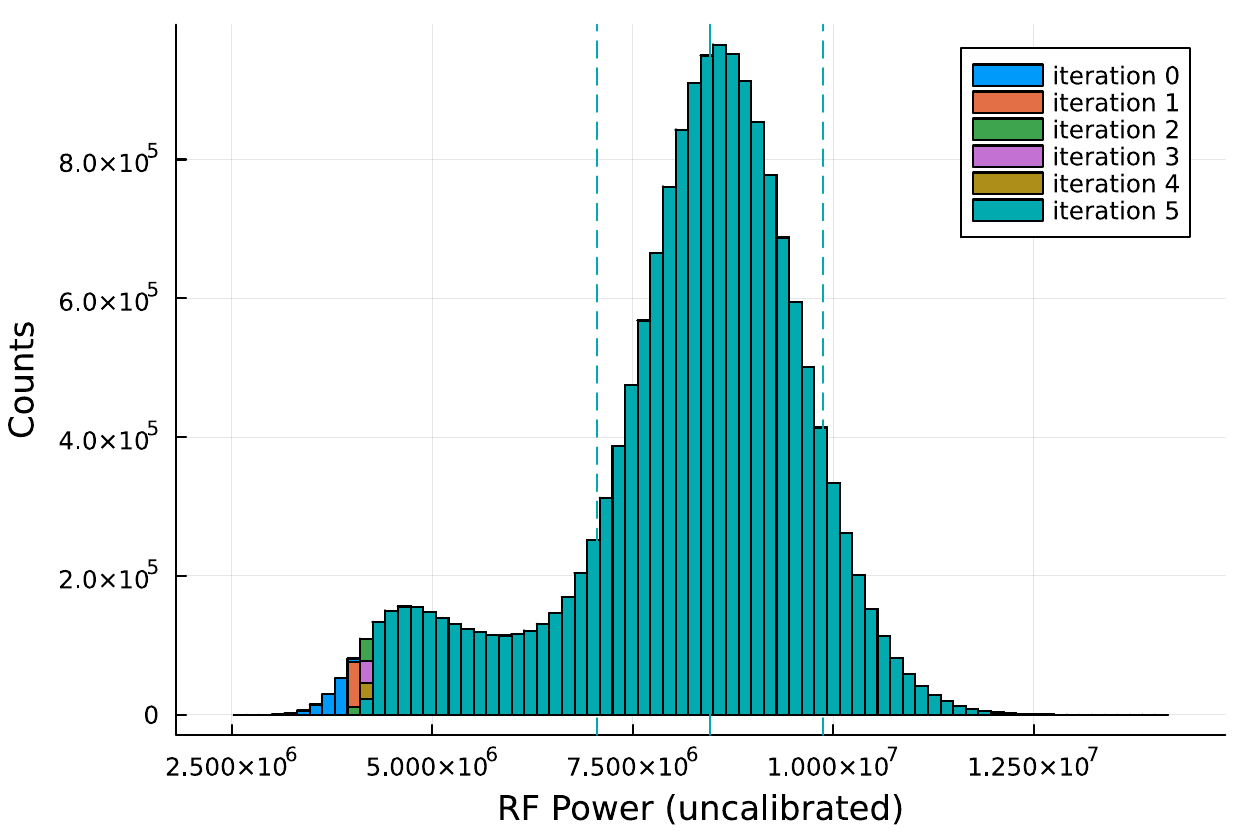}
    \caption{The application of the sigma clipping method to all pixels in the unsummed (nonintegrated), 2D data array for a coarse channel. Before the reduction in variance and "spreading" of the data as a result of time integration, the compactness of the distribution leads to overestimation of the standard deviation attributable to the coarse channel plateau when using a 3$\sigma$ clipping threshold.}
    \label{fig:unsummed_clip}
\end{figure}

When full coarse channels are used for injection, the overlap of the data distributions in the dynamic spectra causes an initial overestimation of the standard deviation, which results in an S/N greater than requested in \texttt{setigen} frames. \texttt{setigen} calculates the intensity of a signal for injection by multiplying a requested S/N by the standard deviation of the full 2D array, then dividing by $N_t^{1/2}$ to counteract the corresponding reduction in background noise and increase in variance of the dataset when $N_t$ time integrations are summed. Due to the overlap of the peaks in Figure \ref{fig:coarse_channel_noise}, the sigma clipping process overestimates the standard deviation in the non-integrated data (see Figure \ref{fig:unsummed_clip}). As the noise of the full array is used for signal injection, this results in injected signals with greater power than intended. When the final spectrogram is time-summed, the sigma-clipping process then correctly calculates the standard deviation of the distribution, and the result is an inflated S/N after normalization.

\subsubsection{Spectral Leakage} \label{subsec_spectral_leakage}

\begin{figure*}
    \centering
    \includegraphics[width=\textwidth]{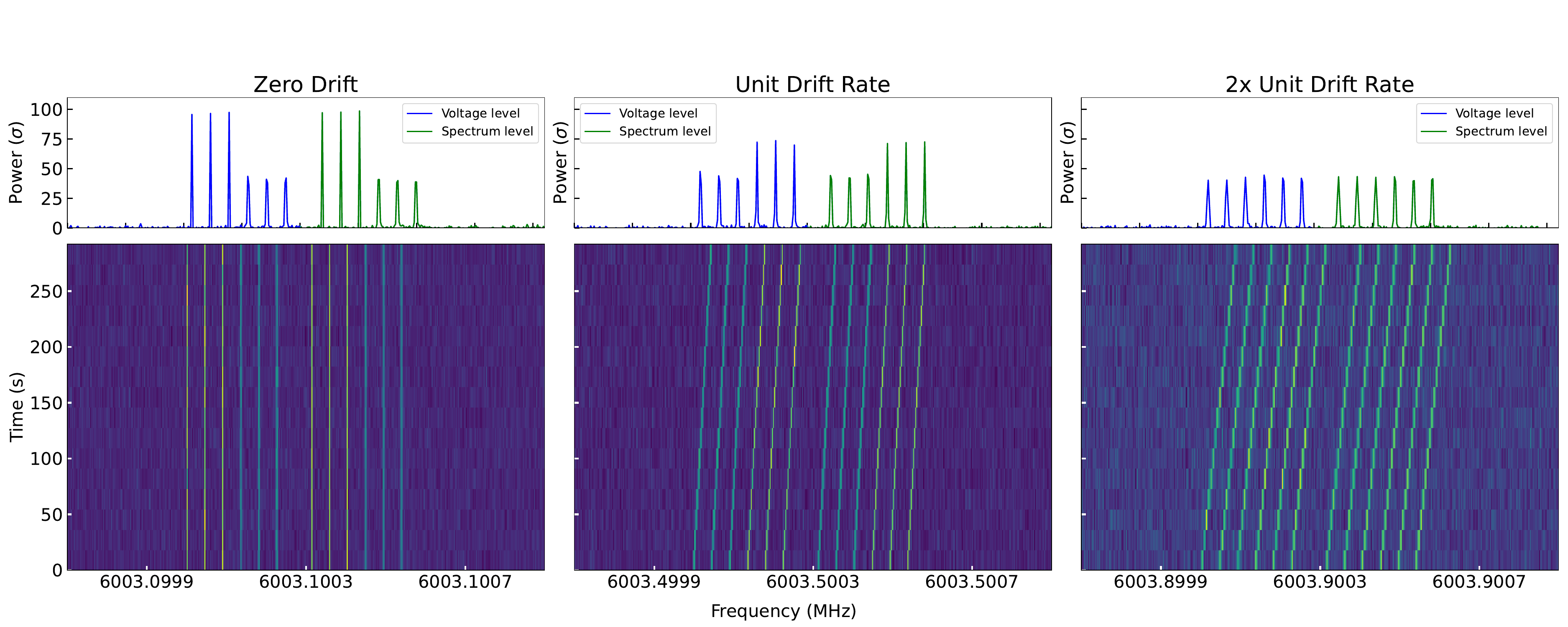}
    \caption{A side-by-side comparison of voltage- and spectrum-level signal injections in synthetic data for bin-centered and offset (edge-centered) signals, stacked with time-summed, dedrifted, normalized power spectra. Each set of six injections consists of three with central frequencies set by fine channel central frequencies (right), and three with central frequencies offset by $\delta f/2$ (left). Sets of six are marked as voltage-level or spectrum-level by color. Full power is recovered in the normalized spectrum for a bin-centered zero-drift signal, and follows Equation \ref{eq:spectral_leakage_attenuation} for offset signals. The power of the drifting signals that result in the integrated spectra is set by the integral of the response (\ref{fig:power_recovered}); here we have chosen $\unit{}$ (unit drift rate) and $2\unit{}$ for clarity.}
    \label{fig:fine_channel_effect}
\end{figure*}

\begin{figure}
    \centering
    \includegraphics[width=\columnwidth]{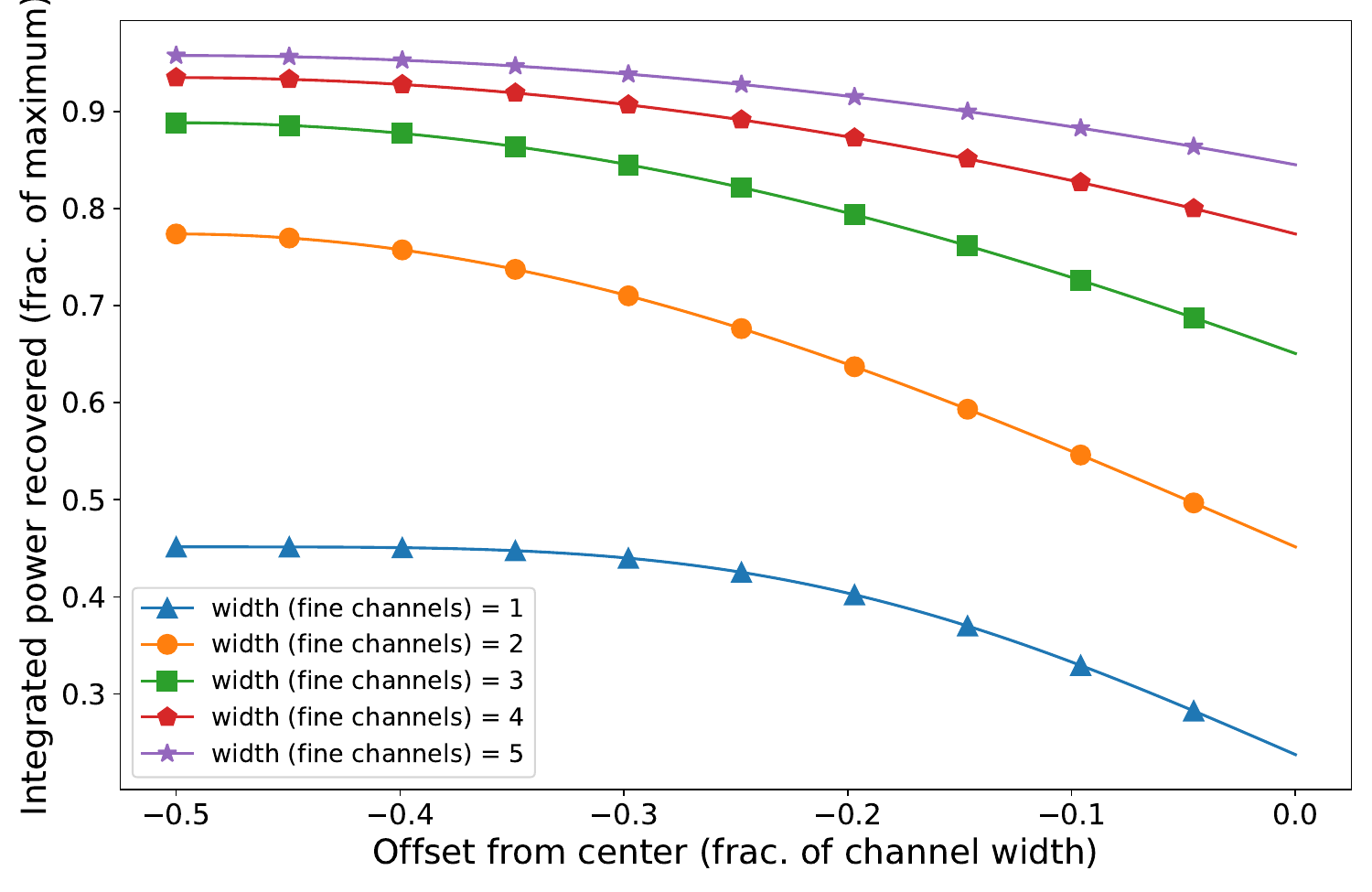}
    \caption{The fraction of maximum integrated power recovered as a function of offset from the center of a frequency bin for various signal bandwidths with a $\text{sinc}^2(f)$ distribution. All curves are computed for a signal drifting at the unit drift rate. Assuming uniform intensity across the bandwidth, signals with broader bandwidths are less susceptible to partial power loss due to fine channel response, as integration over the same bandwidth as a signal drifts through a bin will be higher-valued for broader profile centers.}
    \label{fig:power_recovered}
\end{figure}

In order to observe systematic \tseti{} behavior in regions of sparse or dim RFI, we selected the fine-frequency spectrum of a single node ($N_{coarse} = 64$) from an X-band observation of UGCA\,127 taken on 2021 July 16. Due to reduced RFI density at higher frequencies, \tseti{} detected no signals above a $\tsnr{} = 10$ threshold pre-injection. All signals returned could therefore be clearly identified as resulting from the injection process. We ran several injection tests at high constant S/N and a drift rate of 0\,Hz\,s$^{-1}$ to investigate the frequency dependence of signal injection and assess noise floor levels. We selected central frequencies uniformly spaced across the coarse channel bandwidth in even numbers to avoid interference from the central DC spike, injecting ten signals per coarse channel for a total of 640 injected signals. Again, signals were injected with a $\text{sinc}^2(f)$ frequency profile to simulate the effects of our pipeline's FFT and reproduce the subsequent fine channelization response.

Small variations in recovered power did occur across the bandwidth due to nonuniform noise, but two more significant effects led to the loss of injected signal power. First, \tseti{} version 2.3.2 does not search for 0 drift rate signals; signals inserted with a drift rate of 0 are instead reported with the smallest drift rate \tseti{} searches. The first dedrifting iteration shifts the bottom half of the spectrum by one pixel and results in a factor of two reduction for signals contained within a single frequency bin. Second, the fine channelization process of our data products leads to a variation in recovered intensity at the fine-frequency scale. In our data, the fine channel response is defined by a $\text{sinc}^2(f)$ distribution around the central frequency of the signal. The detected intensity of a non-drifting signal will therefore be attenuated depending on its central frequency with respect to channel edges and the bandwidth of the signal, as

\begin{equation}
\label{eq:spectral_leakage_attenuation}
\textrm{I}_d = \textrm{I}\cdot\textrm{sinc}^2\left(\frac{|\delta f|}{\Delta f}\right),
\end{equation}

\noindent where $I$ is the total power of the signal, $\delta f$ is the frequency offset of the signal's central frequency from the central frequency of the nearest fine channel, $\Delta f$ is the frequency resolution, and $I_d$ is the power that results in data. 

For a linearly drifting signal, power is spread continuously across frequencies, and the apparent recorded intensity becomes an integral of the response across the range of frequencies a signal occupies during a time integration. Figure \ref{fig:fine_channel_effect} details the power reduction experienced by a signal with intensity both constant in time and bandpass profile that occurs as a result of the $\text{sinc}(f)^2$ response in spectral data, depending on whether the signal is centered on a fine channel or its edge. Figure \ref{fig:power_recovered} displays the integrated power recovered as a function of offset and the interaction of offset with signal bandwidth. As we have no \textit{a priori} knowledge of the power with which a signal is sent, understanding these effects and how they present in the power recorded in our data will allow us to more accurately determine the characteristics of a signal and its transmitter in the event of a successful detection, as well as to place more accurate constraints on transmitter populations in the event that no technosignatures are detected.

\subsubsection{Doppler Smearing} \label{subsec_doppler_smearing}

To confirm and quantify \tseti{}'s behavior in real data as a function of drift rate, we again selected ten uniformly spaced central frequencies across each coarse channel bandwidth, aligned with fine channel central frequencies to avoid offset attenuation, for a total of 640 signals. We chose an S/N value of 1000 to capture any sensitivity loss, and varied drift rates across a range of $\pm 5$\,Hz\,s$^{-1}$, wider than our search range, to further test \tseti{} performance with signals that are outside of the user-inputted bounds. Signals were injected with a frequency profile bandwidth of $\sim$5.6\,Hz, equivalent to the width of two fine channels. The \texttt{setigen} \texttt{doppler\_smearing} routine reproduces the effect of Doppler smearing by averaging a given number of copies of a signal between $t$ and $t + 1$, spaced evenly between center frequencies. The resulting ``dechirping loss'' causes attenuation of a signal proportional to its drift, spreading the total power of the signal across adjacent frequency bins in a given integration. 

\begin{figure}
    \centering
    \includegraphics[width=\columnwidth]{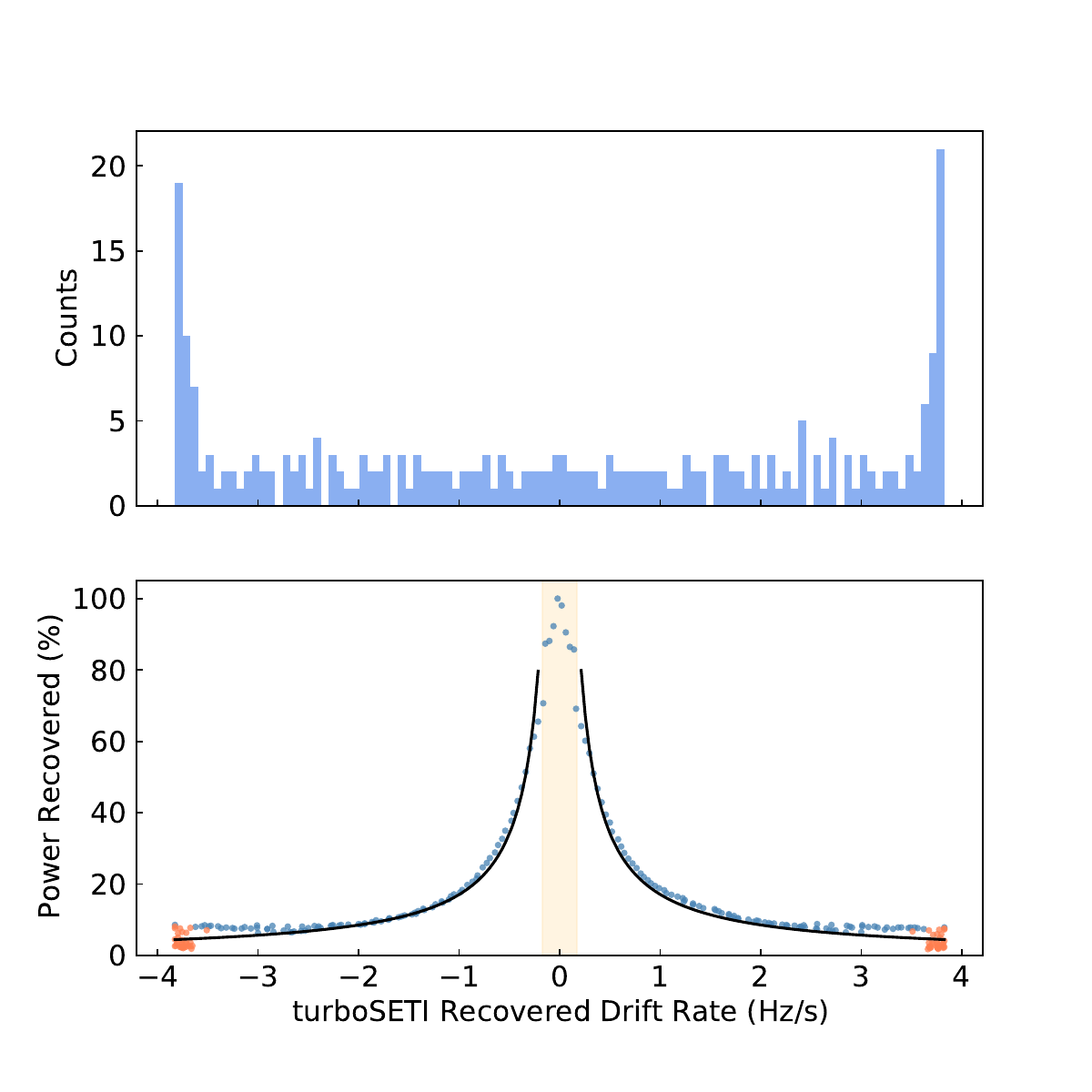}
    \caption{(Top) Distribution of drift rates reported by \tseti{} after injection. Signals were injected with two-channel-wide bandwidths to mitigate response loss (see Figure \ref{fig:power_recovered}. (Bottom) Dechirping efficiency of \tseti{} as a function of Doppler drift rate. The region between $\pm 0.16$\,Hz\,s$^{-1}$ is shaded (orange), and the behavior outside this region is approximated by an $\alpha \frac{1}{N}$ function (black) for both negative and positive drift rates. Dots in orange mark recovered signals with drift rates outside the search bounds.}
    \label{fig:drift_behavior}
\end{figure}

The results of this experiment are shown in Figure \ref{fig:drift_behavior}. Powers are normalized as described in subsection~\ref{noise_calculations}. We find that 100\% of injected signals were recovered with greatly attenuated S/N at high drift rates, reproducing the expected detection efficiency loss as a function of drift rate. We note that, for a signal that is sufficiently smeared, \tseti{} recovers signals with real drift rates greater than its search bounds with inaccurate drift rates and central frequencies. Provided that the power of the signal is sufficient to overcome its attenuation as a result of drift or frequency offset, \tseti{} successfully recovers narrow-band, constant-power signals, including those with drift rates outside its drift bounds. Though we search only a small fraction of drift rate parameter space, this effect demonstrates that the \tseti{} pipeline is capable of detecting some hits outside of its bounds. 

\subsubsection{Limitations and Countermeasures}

The analysis in Sections \ref{subsec_signal_generation}-\ref{subsec_doppler_smearing} calls attention to factors impacting the sensitivity of our search pipeline. Methods to counteract the loss of sensitivity to high-drift signals have been proposed. It is possible to partially mitigate dechirping loss by progressively decreasing the frequency resolution of the dynamic spectra by a factor of two, an approach adopted by \citet{Sheikh_2023}. The ``frequency scrunching" approach allows \tseti{} to operate at near-maximum efficiency by changing the unit drift rate, but also has the potential to produce redundant hits, as each frequency-collapsed dynamic spectrum must be searched with a separate drift rate range. Another approach proposed by \citet{Sheikh_2023} involves coherently correcting raw voltage data before reduction, eliminating sensitivity loss at a specific drift rate but necessitating multiple copies of an observation to cover all drift rates of interest. \citet{Price:2020} proposed to update the \tseti{} algorithm to employ a drift-dependent moving boxcar, averaging $N$ adjacent channels of the integrated spectra to recover smeared power. Future work will implement some of these solutions, and alternative avenues such as machine learning are also being explored \citep{Ma_2023} as well as more computationally intensive overlapping channelization schemes to combat the loss of power due to spectral leakage. At present, no windowing function is applied during the fine channelization of BL data products; doing so could potentially improve channel isolation and reduce spectral leakage. The results presented in this section highlight areas for improvement, and provide useful metrics by which to examine both the results of previous studies and any future technosignature candidate detected using \tseti{}.

We follow the methods of \citet{Enriquez} and past BL papers to calculate EIRP values, using an $\text{N}\sigma$ minimum flux density where the theoretical RMS noise, $\sigma$, is determined using the radiometer equation and the published GBT system equivalent flux densities for each receiver (see Section~\ref{subsec_figs_merit}) and N is the signal-to-noise threshold of our search. In this work, though we ran \tseti{} with $\tsnr{} = 10$, we multiply our threshold by the mode of the factor of difference described in Section~\ref{subsec_figs_merit} when calculating minimum detectable source powers. Our signal injection and recovery analysis indicates that a $\tsnr{} = 10$ threshold for this and past studies conducted using \tseti{} actually corresponds to a detection threshold of 33 times the true noise in a given spectrogram. Detection limits presented in this work account for the corrected threshold. For future injection and recovery projects and technosignature searches, we recommend cropping or correcting for coarse channel shape before running \texttt{setigen} or another injection method. Implementing sigma clipping or another improved noise calculation process would ensure that the desired threshold corresponds to a better measure of spectrogram noise. 

\section{Results} \label{Results}

\begin{figure*}
    \centering
    \includegraphics[width=\textwidth]{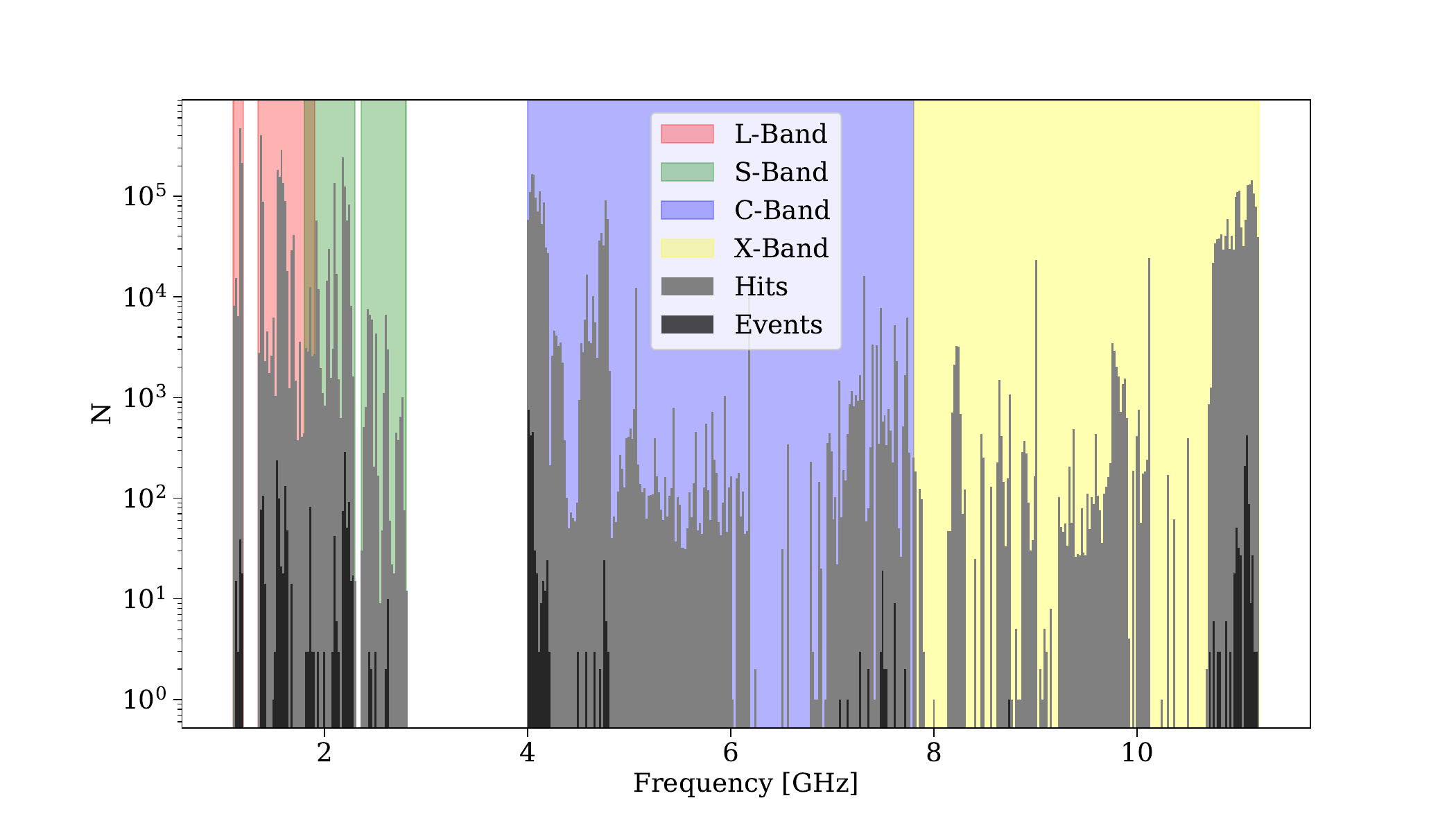}
    \caption{Number of hits (grey) and events (black) versus frequency. The colored regions demarcate the frequency bandwidths of each GBT receiver used in this analysis. We exclude the (unshaded) notch filter regions \citep{lebofsky:19} at L and S bands. Values for each band, with the number of cadences, are shown in Table~\ref{tab:receivers}. Note that there are a different number of cadences at each band, so the number of hits and events plotted here should not be directly compared across bands. The version of \tseti{} used in this analysis does not search for zero-drift signals, so no zero-drift hits are included.}
    \label{fig:freq_hist}
\end{figure*}

\subsection{Signal Distribution} \label{distribution}

\begin{figure*}
    \centering
    \includegraphics[width=2\columnwidth, height=0.88\textheight]{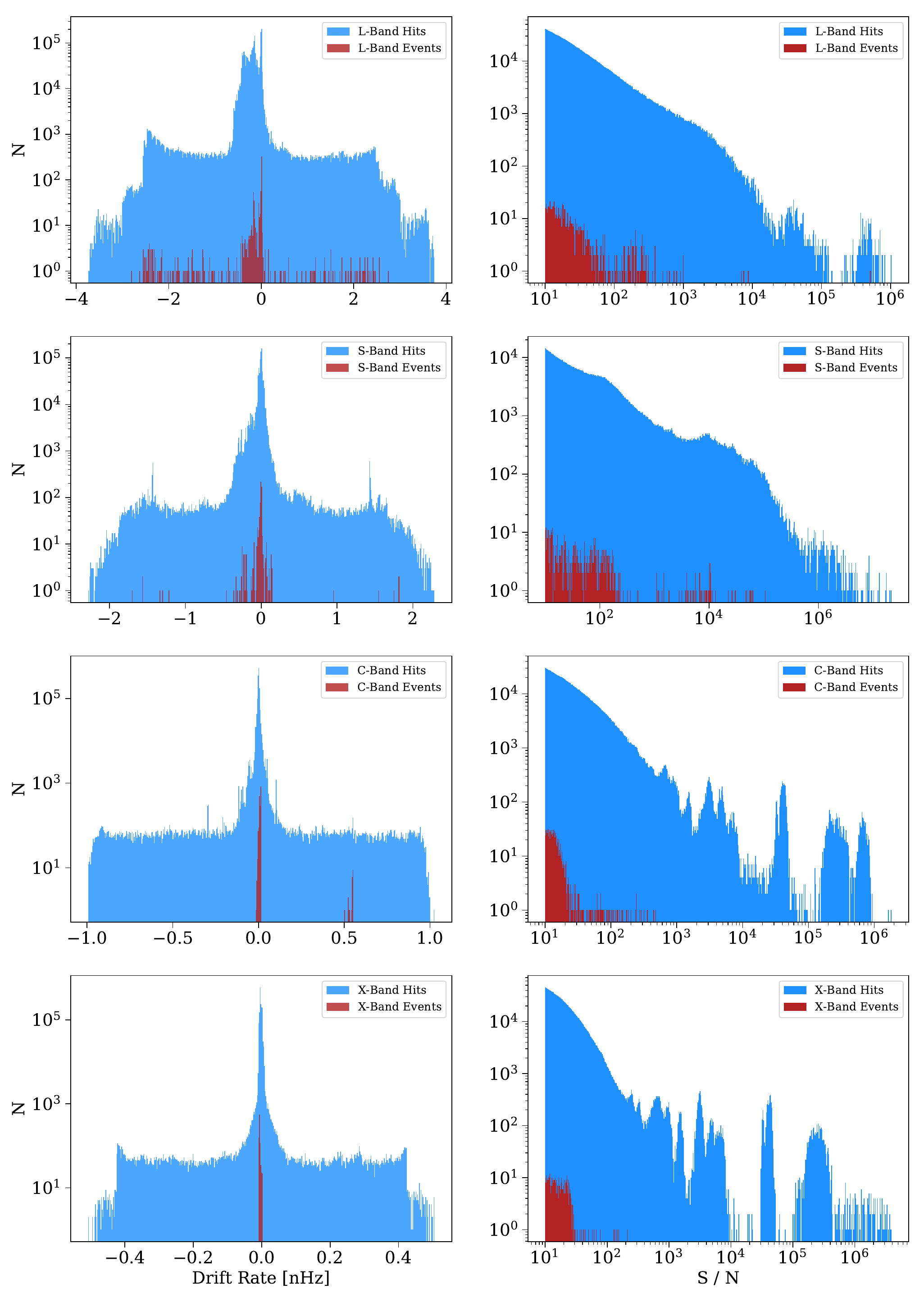}
    \caption{{\em Left Column:} Histograms of drift rates of all hits (blue) and event hits (red) for each GBT receiver for the nearby galaxy sample. Here we normalize the drift rate value by the center frequency of the hit, to produce a value in units of nHz (e.g., 1\,nHz = 1\,Hz\,s$^{-1}$ at 1\,GHz). {\em Right Column:} Histograms of S/N of all hits (blue) and event hits (red) for each GBT receiver. Note that different numbers of cadences were observed at each receiver. Zero-drift hits are not included; the spike at 0 corresponds to the smallest detectable drift rate, our drift rate resolution of $\pm 0.010204$\,Hz\,s$^{-1}$.}
    \label{fig:hit_dist}
\end{figure*}

We ran \tseti{} on 459 cadences, finding a total of 6,034,313 hits and 1,519 events across the four observing bands, distributed as shown in Table~\ref{tab:receivers}. The hit and event frequency distributions are shown in Figure~\ref{fig:freq_hist}. Histograms of the hit distribution as functions of drift rate and S/N are shown in the right and left columns of Figure~\ref{fig:hit_dist}. Although L-band observations only accounted for 21.7\% of the sample, the largest share of hits were found in the L band, at 36.2\%. The C and X bands had similar numbers of hits, at 23.5\% and 26.1\%, respectively, followed by the S band at 14.1\%. Though the S band comprised the largest proportion of cadences and has a historically cluttered RFI environment (see Figure \ref{fig:all_band_spectral_occupancy}), S-band observations contributed the least fraction of both hits and events, with 16.5\% of the total events. The largest fraction of events (42.8\%) was recorded at the C band, double that at the X band (21.8\%) and L band (19.0\%). For all bands, the number of hits falls with S/N, though peaks appear at higher S/Ns which likely indicate populations of RFI transmitters that are either nearby to the telescope or intrinsically bright. While the statistics in Table \ref{tab:receivers} do not include any zero-drift hits, a strong peak still appears at drift rates near zero, as \tseti{} still can recover a signal with bandwidth sufficiently greater than a single fine channel if the intensity is nonuniform.

\subsection{Data Quality and RFI Environment\label{sec:specocc}}

\begin{figure*}
    \centering
    \includegraphics[width=2\columnwidth]{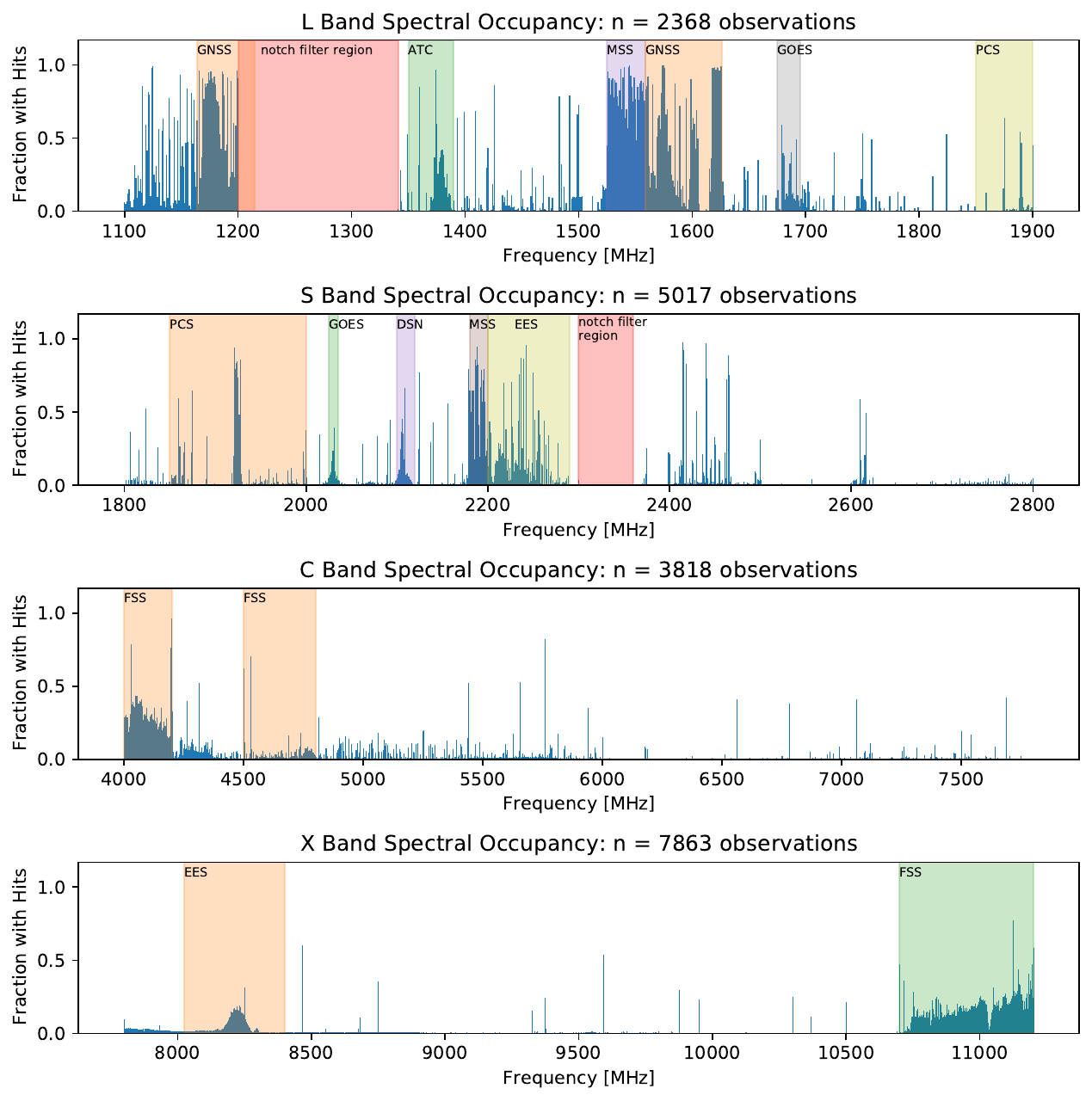}
    \caption{Spectral occupancy for L, S, C, and X band at GBT (from a larger set of data including targets not in the nearby galaxy sample). The red regions in L (1200--1341\,MHz) and S (2300--2360\,MHz) band correspond to the notch filter regions, which are discarded. Notable regions of high spectral occupancy are highlighted and defined in Table \ref{tab:RFI_bands}.}
    \label{fig:all_band_spectral_occupancy}
\end{figure*}

If we assume that all the hits found by \tseti{} are from human-created RFI, the spectral occupancy can be used to assess the quality of the data as a function of frequency. We compute spectral occupancy as the fraction of observations that contain at least one hit in a given 1\,MHz-wide frequency bin. This quantity is calculated using a large ensemble of observations from the BL backend at GBT beginning in 2016, all of which have been passed through \tseti{}. A bin with a spectral occupancy of one means that every observation recorded has at least one hit (and in many cases more than one hit) in that range, and a spectral occupancy of zero would mean that no observation recorded a hit in that range. 

The targets used for the spectral occupancy analysis include some of the observations of the galaxy sample presented in Section \ref{galaxy_sample}, but also archival observations of other targets in order to obtain more precise spectral occupancy statistics. The observations cover L, S, C, and X bands, with 2368, 5017, 3818, and 7863 observations, respectively. Each file corresponds to a single five minute observation spanning the band's observing frequency range as defined in Table \ref{tab:receivers}. The full-sky spectral occupancy plots for L, S, C, and X band are shown in Figure~\ref{fig:all_band_spectral_occupancy}. These plots are similar to plots of received power as a function of frequency that are often used to determine how badly observations may be affected by RFI, but are subtly different given that we are using many observations to calculate the fraction of observations in which signals are detected as a function of frequency. 

\begin{table}
\centering
\caption{Radio Frequency Spectrum Allocations, for bands with high hit densities (see Figure \ref{fig:all_band_spectral_occupancy})}
\begin{tabular}{ll}
\hline \hline
Band & Federal Allocation \\
 (MHz) & \\
\hline
1164-1215 & Global Navigation Satellite System (GNSS) \\
1350-1390 & Air traffic control (ATC) \\ 
1525-1535 & Mobile-satellite service (MSS) \\
1535-1559 & MSS \\
1559-1626.5 & GNSS \\
1675-1695 & Geostationary operational \\ & environmental satellite (GOES) \\
1850-2000 & Personal communications services (PCS) \\
2025-2035 & GOES \\
2100-2120 & NASA Deep Space Network use (DSN) \\
2180-2200 & MSS \\
2200-2290 & Earth exploration satellite (EES) \\ 
4000-4200 & Fixed Satellite Service (FSS) \\
4500-4800 & FSS \\ 
8025-8400 & EES \\
10700-11200 & FSS\\
\hline
\end{tabular}
\begin{tablenotes}
\item  Modified and extended from \cite{Price:2020},
\item FCC  Table of Frequency Allocations\footnote{https://transition.fcc.gov/oet/spectrum/table/fcctable.pdf}
\end{tablenotes}
\label{tab:RFI_bands}
\end{table}

\begin{figure*}
\centering
\subfigure[C band hit density heatmap of the southern half of the sky]{\label{subfig:cband_north}\includegraphics[width=2\columnwidth]{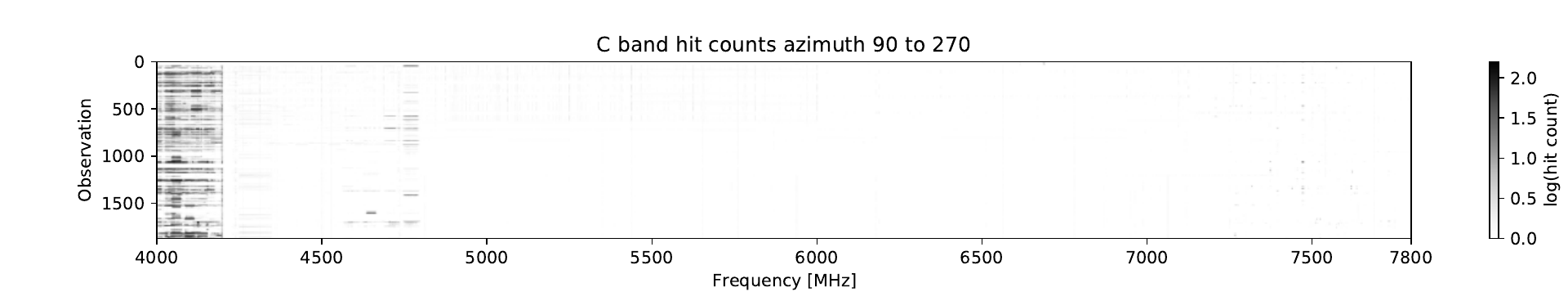}}\\ 
\subfigure[C band hit density heatmap of the northern half of the sky]{\label{subfig:cband_south}\includegraphics[width=2\columnwidth]{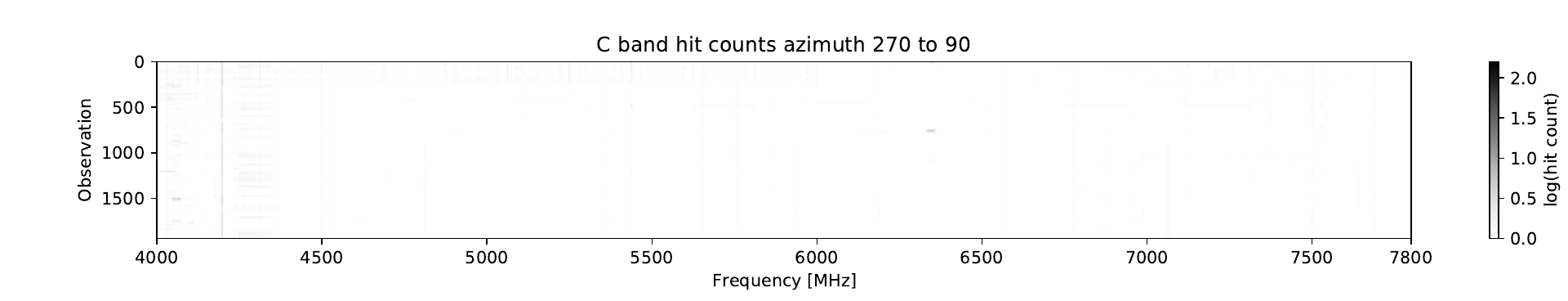}} 
\caption{C band \tseti{} hit counts (from the larger set of observations described in \S~\ref{sec:specocc}) divided up into northern (bottom panel) and southern (top panel) halves of the sky. Each horizontal line in these plots represents the number of hits detected per 1\,MHz bin in a given observation. The thousands of observations plotted here were taken over the course of several years of the BL program at GBT. Noticeable differences in the hit distributions include increased hit counts in the range of 4000-4200\,MHz which appear when the telescope was pointed towards the southern half of the sky and absent when pointed towards the north.}
\label{fig:cband_north_south_heatmap}
\end{figure*}
\begin{figure*}
\centering
\subfigure[X band hit density heatmap of the southern half of the sky]{\label{subfig:xband_north}\includegraphics[width=2\columnwidth]{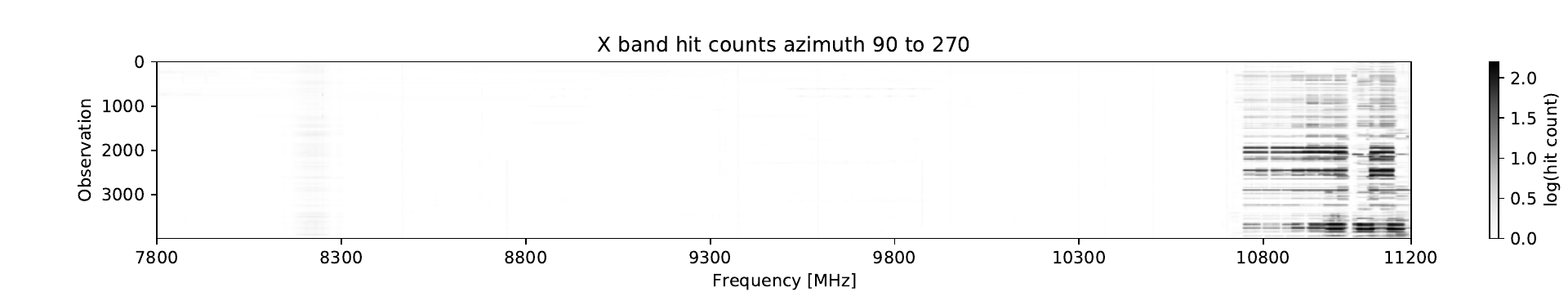} } 
\subfigure[X band hit density heatmap of the northern half of the sky]{\label{subfig:xband_south}\includegraphics[width=2\columnwidth]{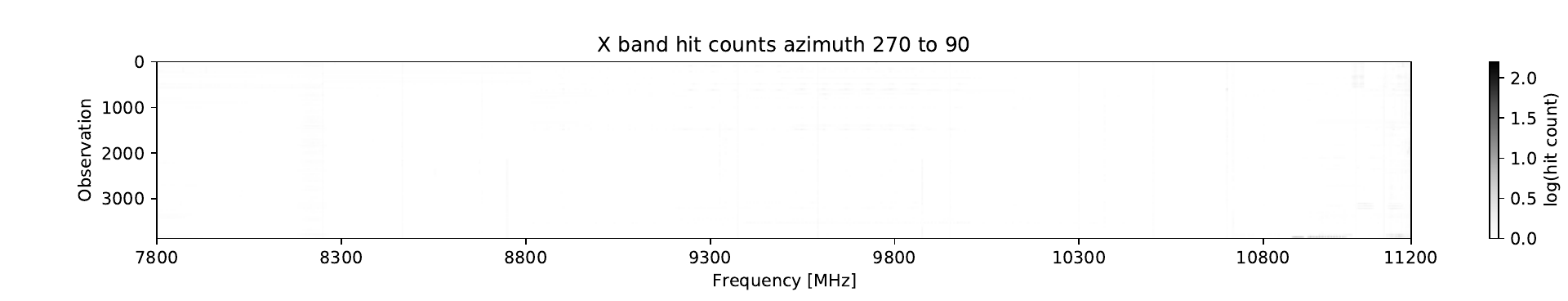}} 
\caption{X band \tseti{} hit counts (from the larger set of observations) divided up into northern (bottom panel) and southern (top panel) halves of the sky. Noticeable differences in the hit distributions include increased hit counts around 8100-8300\,MHz and 10700-11200\,MHz, both of which appear when the telescope was pointed towards the southern half of the sky and absent when pointed towards the north.}
\label{fig:xband_north_south_heatmap}
\end{figure*}

The observation data were grouped by various parameters to qualitatively search for patterns in the hit counts, including grouping by telescope altitude and azimuth and the local time of observation at the GBT. The most apparent differences appeared when splitting the azimuth angle into northern and southern halves of the sky. The southern sky exhibited higher hit counts in the C band frequency ranges of 4000--4200\,MHz, caused by the top of the C band geostationary satellite downlink range, (see Figure \ref{fig:cband_north_south_heatmap}) and X band frequency ranges of 8100--8300\,MHz, and 10700--11200\,MHz, (see Figure \ref{fig:xband_north_south_heatmap}). In all three cases, the increased hit counts appear in bands with known satellite transmissions, given in Table \ref{tab:RFI_bands}.

\subsection{Assessing the False Positive Rate}

\begin{figure*}
    \centering
    \includegraphics[width=2\columnwidth]{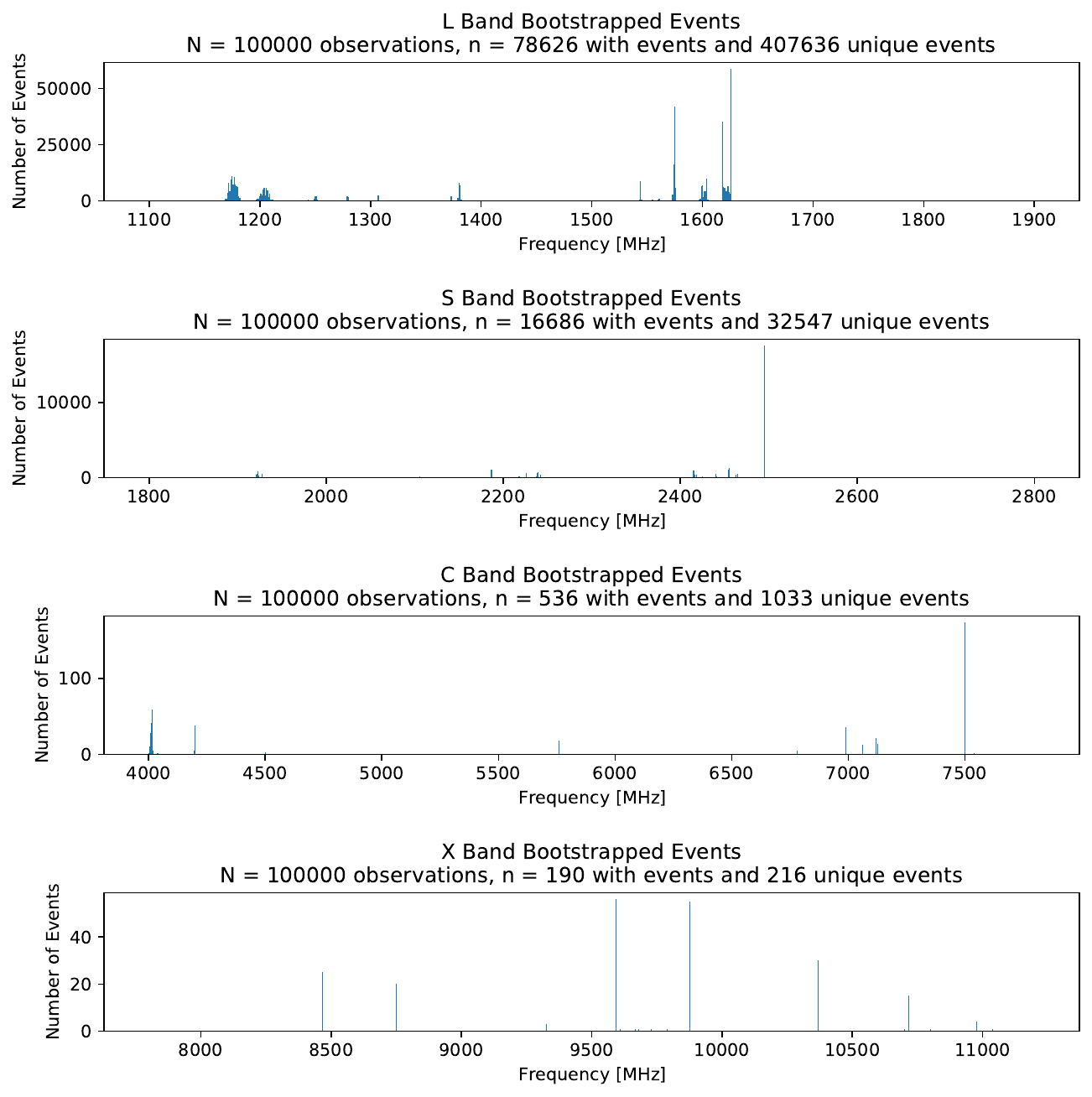}
    \caption{Distribution of bootstrapped event detections (for the sample described in \S~\ref{sec:specocc}) as a function of frequency. \textbf{L band}: the highest number of events, 58,618 was detected in the 1626\,MHz bin. \textbf{S band}: the highest number of events, 17,520 was detected in the 2495\,MHz bin. \textbf{C band}: the highest number of events, 173 was detected in the 7500\,MHz bin. \textbf{X band}: the highest number of events, 56 was detected in the 9591\,MHz bin.}
    \label{fig:bootstrap_events}
\end{figure*}

\begin{figure*}
    \centering
    \includegraphics[width=2\columnwidth]{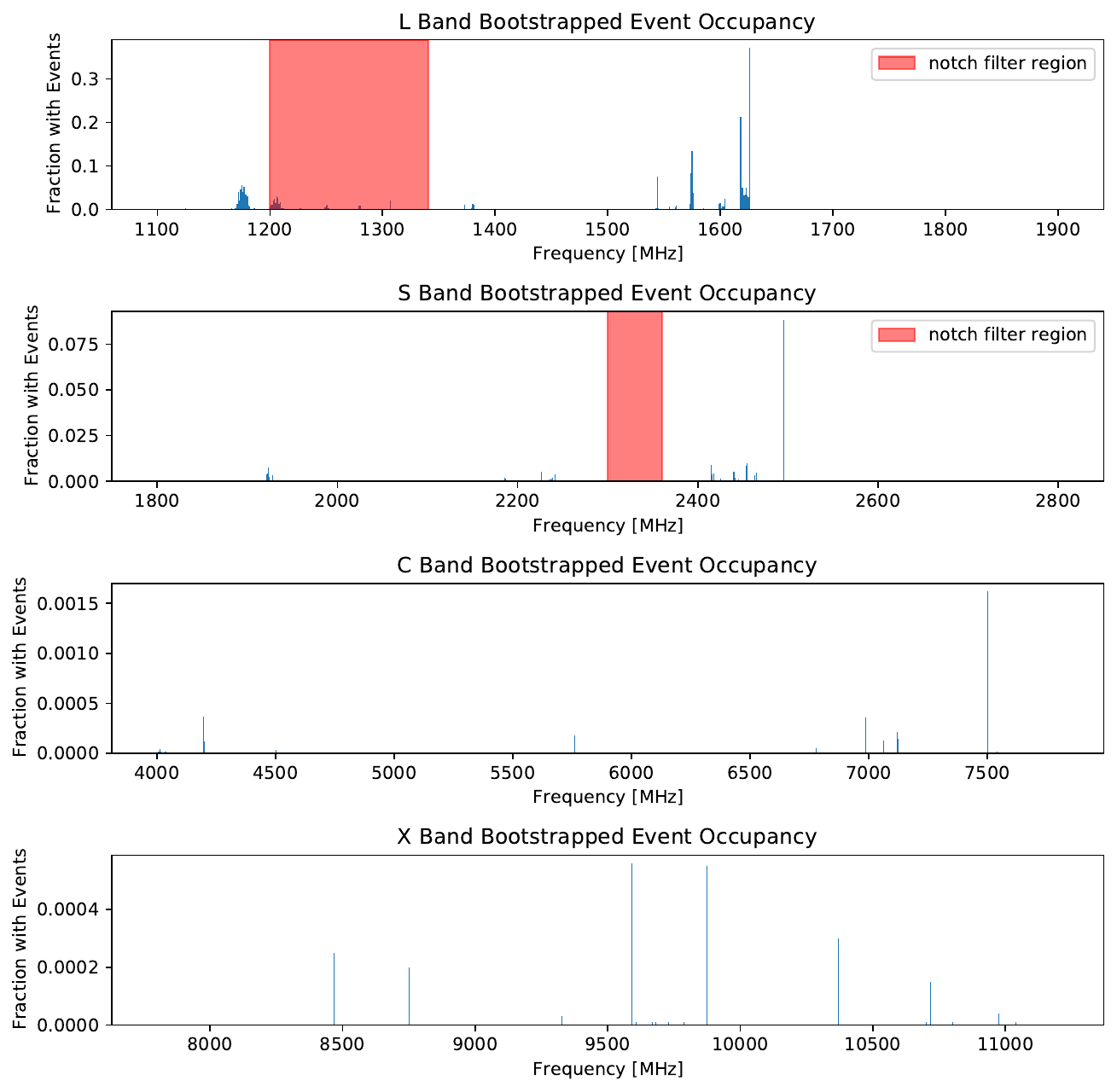}
    \caption{Distribution of bootstrapped event occupancy -- the fraction of simulated observations in which at least one event was detected -- as a function of frequency. \textbf{L band}: The 1626\,MHz frequency bin had the highest occupancy, with 37\% of simulated observations detecting at least one event. \textbf{S band}: The 2495\,MHz frequency bin had the highest occupancy, with 8.8\% of simulated observations detecting at least one event. \textbf{C band}: The 7500\,MHz frequency bin had the highest occupancy, with 0.16\% of simulated observations detecting at least one event. \textbf{X band}: The 9591\,MHz frequency bin had the highest occupancy, with 0.056\% of simulated observations detecting at least one event.}
    \label{fig:bootstrap_occupancy}
\end{figure*}

Intermittent RFI can lead to false positive events when it appears, by random chance, in three consecutive ON observations and is absent in the intervening OFF observations. To estimate how often this occurs as a function of frequency, we bootstrapped the individual observations used in the TESS target search of \cite{Traas}, and re-ran \texttt{FindEvent} to create artificial events. This was done by randomly selecting six files (three on and three off targets) without replacement. In general, the observation cadence is now ABCDEF, instead of the usual ABACAD, and unlike for the standard ABACAD cadences, the targets were not necessarily observed consecutively in time. The list of files are then passed into \tseti{}’s find event pipeline with modified timestamps to appear as if the files all came from a single observation cadence. As we do not expect a technosignature candidate to be made up of hits coming from different positions in the sky at different times, we can be confident that any events found are false positives due to intermittent RFI. We performed 100,000 random draws for each observing band. Additionally, the event occupancy was also calculated. The process for calculating the event occupancy is identical to that of spectral occupancy, with the key difference being that only events are counted, as opposed to all hits detected during an observation. The histogram showing the event distributions can be found in Figure \ref{fig:bootstrap_events} and the event occupancy in Figure \ref{fig:bootstrap_occupancy}. It can be seen that C and X band had the fewest events resulting from chance, with 1033 and 216 events in 100,000 simulated observations. L and S band had a much higher number of events with 407,636 and 32,547 events, respectively. 

The frequency of chance events at L and S band is higher than what is predicted by Equation \ref{eq:p_event} --- 37\% of observations record an event in the 1626\,MHz bin and 8.8\% of observations record an event in the 2595\,MHz bin. These can be attributed to the high density of radio transmissions at these frequencies, which is smoothed over during the spectral occupancy calculation. The spectral occupancy calculation treats any number of hits greater than zero exactly the same, whether there is one narrowband signal in the bin or many narrowband signals packed densely as seen in the examples shown in Appendix \ref{app:rfi_context}. With the higher number of hits per bin, the detection of a higher number of events can be expected, resulting in an increased event occupancy. 

\subsection{Event Ranking}

The spectral occupancy gives us the probability of observing a hit in a given frequency bin. Thus for an event in an ABACAD cadence, we multiply the probability of observing a hit in the ``on” targets with the probability of not observing a hit in the ``off” targets. We find that the probability of an event is given by 

\begin{equation}
\label{eq:p_event}
P_{\textrm{event}}(o(f)) = o(f)^3(1 - o(f))^3
\end{equation}

\noindent where o(f) is the spectral occupancy for a given frequency bin. As a result, we find that we are most likely to find events at frequencies with a spectral occupancy of 50\%, at which a predicted maximum of 1.56\% of observations yield events at that frequency. As the spectral occupancy moves away from 50\%, the likelihood of an event decreases, with events being least likely at frequencies with spectral occupancies near 100\% or 0\%. 

With the prevalence of RFI and the frequency with which it gets flagged as a potential extraterrestrial technosignature \citep{Enriquez, Price:2020, Traas, Franz_2022}, we want a way to tell us which events to inspect first. 
Equation \ref{eq:p_event} allows us to use the spectral occupancy to quantify the likelihood of detecting an event, but given the number of events that are detected, it would be useful to have a way to quantify which events should be looked at first. One method is to look at the events that are least likely to occur, specifically at frequencies that have a low spectral occupancy. This can be achieved by adding a bias term to Equation \ref{eq:p_event} to give a higher weight to frequencies with low spectral occupancy and a lower weight to those with higher spectral occupancy. Here we divide Equation \ref{eq:p_event} by a cubic bias term, resulting in 

\begin{equation}
    P_{\textrm{weighted}}(o(f)) = (1 - o(f))^3
    \label{eq:p_weight}
\end{equation}

\noindent where $o(f)$ is the spectral occupancy for a given frequency bin. This can be interpreted as the uniqueness of the candidate based on the likelihood of a signal not appearing in the OFF observations. For example, an event appearing in a frequency bin with a spectral occupancy of zero would be weighted as most interesting, while an event appearing in a bin with a spectral occupancy of one would be weighted as least interesting. 

\begin{figure*}
    \centering
    \includegraphics[width=2\columnwidth]{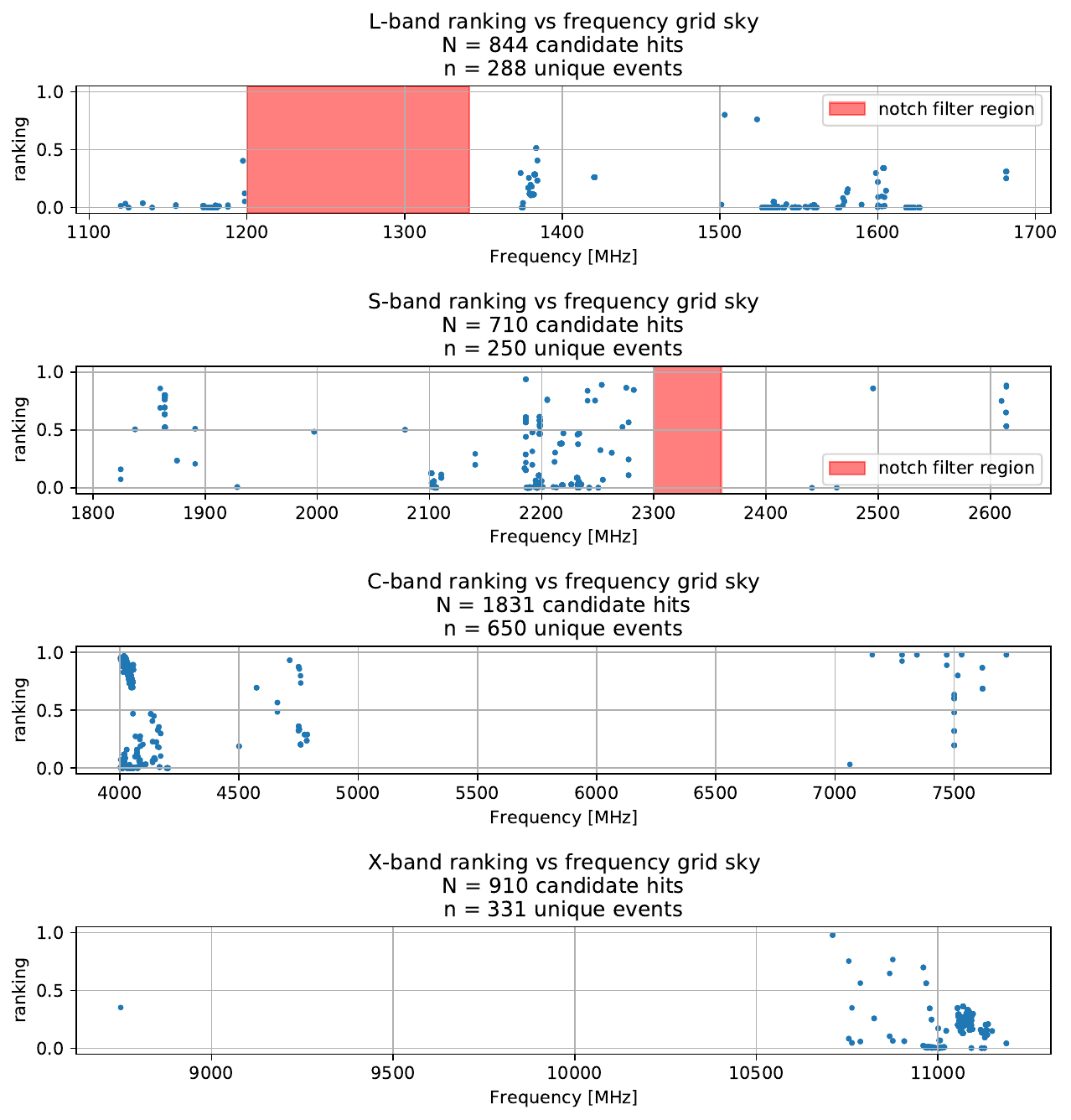}
    \caption{\textbf{L band}: The highest ranked event occurs at a frequency of 1502.993574\,MHz, and was given a ranking of 0.801, which corresponds to a spectral occupancy of 0.071. \textbf{S band}: The highest ranked event occurs at a frequency of 2185.955996\,MHz, and was given a ranking of 0.94, which corresponds to a spectral occupancy of 0.021. \textbf{C band}: The highest ranked event occurs at a frequency of 7531.252936\,MHz, and was given a ranking of 0.986, which corresponds to a spectral occupancy of 0.0047. \textbf{X band}: The highest ranked event occurs at a frequency of 10710.079523\,MHz, and was given a ranking of 0.979, which corresponds to a spectral occupancy of 0.072.}
    \label{fig:event_ranking}
\end{figure*}

With the qualitative difference in the hit distributions based on position in the sky such as those seen in Figures \ref{fig:cband_north_south_heatmap} and \ref{fig:xband_north_south_heatmap}, we divided the sky into three altitude bins and four azimuthal bins, corresponding to widths of $30\degr$ and $90\degr$, respectively, and calculated the spectral occupancy for each region independently. This has the advantage of grouping hits that are nearby on the sky, increasing their influence on the ranking of nearby events and reducing their impact on events in a different region. These grouped spectral occupancies have been applied to the 1,519 events detected in the nearby galaxy survey. The distribution of events can be seen plotted against frequency and their ranking under Equation \ref{eq:p_weight} in Figure \ref{fig:event_ranking}. 

\section{Event Grouping and Rejection}

Vetting of events was primarily conducted visually, using plots generated with \texttt{plot\_event\_pipeline}. To avoid missing a signal, all 1,519 events found in the nearby galaxy sample were inspected by eye. However, for comparison with spectral occupancy statistics and to improve RFI characterization, we also grouped events using the \texttt{sklearn} DBSCAN clustering algorithm \citep{DBSCAN}. DBSCAN deals well with outlier detection in noisy, few-parameter datasets. Here, its method of calculating Euclidian distances between scaled data points is effective at assembling populations of events at similar frequencies and drift rates. Grouping was conducted after masking frequencies outside the bands shown in Table \ref{tab:receivers} as well as within the GBT L-band and S-band notch filters. Event groups that align with high-incidence regions of spectral occupancy are more likely to correspond to ensembles of persistent known interferers than large collections of ETI beacons. Therefore we can assess event groups with the localization criterion as per Section \ref{distribution} and add the context of correlation with regions of high spectral occupancy and known RFI allocations. We averaged drift and frequency values across each set of three event hits, clustered in frequency and drift, then ranked outliers according to Euclidean distance from the nearest cluster in parameter space. We chose a value of $\epsilon$ = 0.075 for the maximum neighborhood radius using the nearest-neighbors ``elbow method" for DBSCAN clustering \citep{RahmahSittanggang:2016}. As we clustered in frequency and drift rate, we followed the methods of \citet{DBSCAN} for two-dimensional datasets and chose the minimum number of points per cluster to be twice the dimensionality of our dataset, $N_{points} = 4$.

\begin{table}
\centering
\caption{Event Cluster Characteristics}
\begin{tabular}{rrlll}
Cluster & Events & Drift Range & Frequency Range & Band \\
&& (Hz s$^{-1}$) & (MHz) & \\
\hline
1 & 363 & {[}-0.17, 0.11{]} & {[}1120.0, 2463.1{]} & L, S \\
2 & 5 & {[}-0.66, -0.60{]} & {[}1374.3, 1600.3{]} & L \\
3 & 55 & {[}-0.30, -0.15{]} & {[}1134.0, 1618.1{]} & L \\
4 & 5 & {[}-1.48, -1.43{]} & {[}1420.6, 1626.4{]} & L \\
5 & 11 & {[}-0.49, -0.38{]} & {[}1575.7, 1622.2{]} & L \\
6 & 4 & {[}-1.31, -1.24{]} & {[}1618.2, 1626.1{]} & L \\
7 & 12 & {[}-0.52, -0.47{]} & {[}2216.4, 2219.4{]} & S \\
8 & 6 & {[}0.14, 0.16{]} & {[}2231.7, 2233.4{]} & S \\
9 & 5 & {[}-0.26, 0.20{]} & {[}2233.1, 2277.5{]} & S \\
10 & 5 & {[}-0.01, 0.01{]} & {[}2610.0, 2614.3{]} & S \\
11 & 612 & {[}-0.03, 0.06{]} & {[}4003.7, 4202.5{]} & C \\
12 & 14 & {[}-0.01, 0.01{]} & {[}4500.1, 4786.0{]} & C \\
13 & 10 & {[}-0.03, 0.01{]} & {[}7062.5, 7718.8{]} & C \\
14 & 9 & {[}4.07, 4.13{]} & {[}7500.0, 7500.1{]} & C \\
15 & 329 & {[}-0.08, -0.01{]} & {[}10710.1, 11188.3{]} & X
\\ \hline
\end{tabular}%
\label{tab:clusters}
\end{table}

Clustering resulted in 15 distinct clusters and 71 events flagged as outliers. The bounds and number of events assigned to each cluster are reported in Table \ref{tab:clusters}. Three clusters (1, 11, and 15) comprised the vast majority of events, with cluster 11 spanning the region from 4000--4200\,MHz accounting for a full 58\% of events. As this frequency range corresponds with a region of high spectral occupancy and an FCC allocation for geosynchronous satellites, and visual inspection shows many subgroups with similar modulation, we attribute those events to RFI. The high rate of false positives resulting from this small range indicate that it may be valuable to time observations to avoid the geostationary belt. Other clusters occur at high-spectral occupancy regions and correspond to known RFI allocations; clusters 7-10 fall within the 2200--2290\,MHz band used for spacecraft tracking and telemetry, cluster 15 also aligns closely with a fixed-satellite communications allocation in X-band, and cluster 1 spans several navigational and telecommunications allocations (see \citealt{Price:2020}) with a distribution of drift rates close to zero. Cluster 1 represents a pitfall of clustering with a small number of parameters that have significant range; Cluster 1 spans both L and S band due to the close grouping of drift rates, despite frequency interruptions such as the notch filter and separation between bands. Inclusion of additional hit characteristics as clustering parameters, such as kurtosis and bandwidth, could improve the accuracy and information density of this clustering approach. Future work will pursue clustering in a higher-dimensional space, as well as expanded techniques in anomaly detection for the classification of RFI populations.

Outliers were re-inspected for anomalous structure. Examples of the top-ranked outliers are shown in Appendix~\ref{app:outliers}. The majority of outlier events are stragglers of other clusters that can be identified as part of easily identifiable clusters of RFI, such as the Iridium satellite constellation and the aeronautical navigation allocation identified with cluster 7. Occasional high-drift features are flagged as events due to wideband, nonlinear RFI. Two C-band outliers, at 7499\,MHz and 7514\,MHz, respectively, appear in ``on'' and ''off'' scans and are likely associated with meteorological satellites (e.g., RosHydroMet) transmitting at $\sim$7500\,MHz, alongside cluster 14. 

Many signals appear visually in both the ON and OFF scans of a cadence, but were not detected by the pipeline because their S/Ns were below the threshold in the off scans. This common pattern likely indicates a source of interference local to the telescope itself, and could potentially be mitigated by running the pipeline with a lower S/N threshold for off scans and checking for continuity before discarding hits as described in Section \ref{Doppler Search}. Ultimately, we determined that all events were attributable to RFI.

\section{Discussion}

\subsection{Figures of Merit} \label{subsec_figs_merit}

The large and complex parameter space of SETI searches makes it difficult to quantify the comprehensiveness of a study. Efforts to examine the rigor of large surveys with respect to certain parameters or subsets of parameters have resulted in the development of various ``figures of merit''. Here, we calculate the most widely used figures of merit, comment on their caveats, and provide values from other similar surveys to facilitate comparison. 

The Drake Figure of Merit \cite[DFM;][]{Drake:1984} is a well-known metric that endeavors to allow comparisons of physical survey characteristics. It is given by

\begin{equation}\label{eq:DFM}
    \textrm{DFM} = \frac{n~\Delta \nu_{\textrm{tot}}~\Omega}{F^{3/2}_{\textrm{min}}},
\end{equation}

\noindent where $n$ is the number of sky pointings, $\Delta \nu_{\textrm{tot}}$ is the total observing bandwidth, $\Omega$ is the full width at half maximum of the receiver, and $F_\textrm{min}$ is the minimum detectable flux density. As the DFM depends on antenna characteristics, it must be calculated separately for different receivers; surveys using multiple detectors may compute a combined figure as the sum of the DFM at each. All else being equal, larger figures indicate more comprehensive surveys. For our survey, we found the DFM to be $\sim$ $1.71\times10^{32}$ for signals with drift rates $\le$ 0.16\,Hz\,s$^{-1}$, and $\sim$ $1.46\times10^{30}$ for the highest drift rates we searched.

The DFM is a useful heuristic for comparing the breadth of searches in frequency and geometric space, but as discussed by \citet{Price:2020} and \citet{Margot-2021}, its focus on sky coverage and minimum detectable flux density neglect potential nuances of ETI transmitter distributions and stellar density as well as other dimensions of survey sensitivity. \citet{Enriquez} define one approach to an improved heuristic: the Continuous Waveform Transmitter Rate Figure of Merit (CWTFM), defined as 

\begin{equation}\label{eq:CWTFM}
    \textrm{CWTFM} = \zeta_{\textrm{AO}} ~ \frac{\textrm{EIRP}_{\textrm{min}}}{N_\textrm{stars}~\nu_{\textrm{rel}}}
\end{equation}

\noindent where $\nu_{\text{rel}}$ is the fractional bandwidth, $N_{\text{stars}}$ is the number of pointings times the number of stars per pointing, and the $EIRP_{\text{min}}$ is the minimum detectable equivalent isotropic radiated power (EIRP) in units of Watts.  $\zeta_{\text{AO}}$ is a normalization constant such that the CWTFM is 1 for an EIRP equal to that of Arecibo. $EIRP_{\text{min}}$ depends on the distance to the target as well as the minimum detectable flux density, and denotes the power requirements on a putative transmitter located in the most distant source in our sample. 

\begin{figure*}
    \centering
    \includegraphics[width=2\columnwidth]{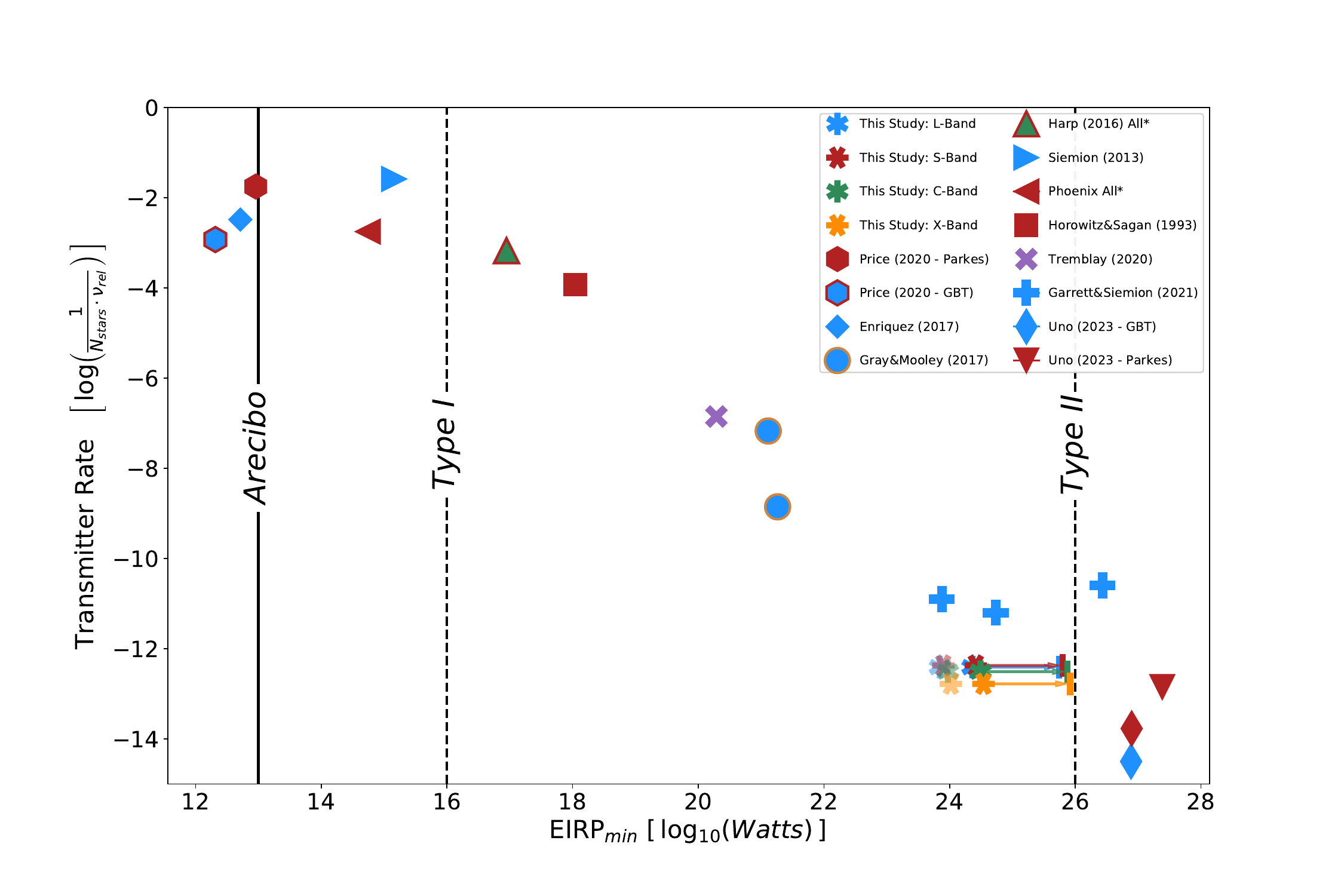}
    \caption{A comparison of transmitter rate versus EIRP$_{\textrm{min}}$ for this work with previous searches, where blue, red, green, and yellow correspond to searches of the L, S, C, and X bands respectively. The three vertical lines mark the EIRP of the Arecibo S-Band radar, the power level of a Kardashev Type I civilization, and the total power budget of a Kardashev Type II civilization (i.e., the total solar luminosity) \citep{sagan1975cosmic}. Only the results for this work, here shown as six-pointed stars, are corrected for \tseti{} noise calculation methods and calculated using an S/N of 33 rather than an S/N of 10. We note that the results of all other studies conducted with \tseti{} \citep{Uno_2023, GarrettSiemion:2022, Enriquez, Price:2020} also represent searches off by a factor of $\sim 3.3$, but we have elected not to alter them here. We show the lower bounds of the EIRP limit that results from the variation in dechirping efficiency extending from zero drift rate to the maximum drift rate searched. The transparent six-pointed stars are included for comparison with past works, and represent the $\text{EIRP}_{min}$ for this work if we take the parameter \tsnr{} = 10 at face value as a $10\sigma$ limit (in other words, with no sensitivity correction for \tseti's noise calculation).}
    \label{fig:eirp}
\end{figure*}

Figure~\ref{fig:eirp} shows the EIRP plotted against the transmitter rate (defined as $1/N_{\text{stars}}~\nu_{\text{rel}}$) alongside values from other surveys for comparison. Figures of merit were calculated using only unique pointings; that is, counting each galaxy once. Estimates (Appendix~\ref{app:galaxy_properties}) for the number of stars observed (to determine the transmitter rate) were derived from K-band luminosities collated by \citet{isaacson:2017}, assuming the average mass of a star is $1 M_\sun$, and calculating the percentage of each galaxy covered by a GBT pointing at each band assuming a uniform stellar density across the galaxy's area. Technosignature searches are compromises between sensitivity (higher sensitivity towards the left-hand side of the figure) and sky and bandwidth coverage (more stars, and/or wider fractional bandwidth coverage, towards the bottom of the figure).

\subsection{Transmitter Limit}

Given our lack of a technosignature detection, we can calculate the transmitter limit, or the maximum percentage of nearby galaxies that possess a transmitter detectable with our system and search parameters. \cite{Price:2020} and \cite{Traas} calculate this limit using a one sided 95\% Poisson confidence interval with a 50\% probability of actually observing a signal if the transmitter is present \citep{1986ApJ...303..336G}, interpreted as a 97\% Poisson confidence interval. Calculated in the same manner for comparison, using the tables of \cite{Gehrels}, we find an upper limit of 3.8\% at all four bands. However, we offer an alternative method for determining upper limits on transmitter rates that we believe to be more faithful to a frequentist interpretation. For a Poissonian model, the cumulative distribution function gives the probability of detecting $\text{N}_d = \text{n}$ events or fewer, assuming some underlying model is true. To find the upper limit on the the transmitting fraction $\text{f}_{tr}$, we must exclude all models such that

\begin{equation}
    \text{CDF}(n;\text{f}_{tr}) \le 1 - p
    \label{eq:CDF}
\end{equation}

\noindent where $p$ is the desired confidence level. 

For our analysis, we choose a one-sided 95\% confidence interval, corresponding to a $p$ value of 0.95. Since we detected $\text{n} = 0$ events, Equation \ref{eq:CDF} gives an $\text{N}_d$ value of 3 for a Poisson distribution \citep{Gehrels}. We count each galaxy once for a total of 97 targets, so the maximum fraction of galaxies in our sample which could possess a transmitter of sufficient power is $f_{tr} = 3/97 = 3.1\%$. With the caveats discussed in Section \ref{Doppler Search} and Section \ref{Results}, we report that a lack of a successful detection at any of 97 distinct targets results in an upper limit of 3.09\%  of targets at all four bands (L, S, C, X), for narrowband, 100\% duty cycle transmitters with equivalent isotropic radiated power on the order of $10^{24}$\,Watts.

\subsection{Limits on Repeatability}

Radio emission phenomena traveling significant distances in intergalactic and interstellar space are necessarily subject to the effects of scattering and the fluctuating density of free electrons along their paths. Scattering can produce a number of observable phenomena, but most significant to a narrowband search are spectral broadening, which can spread signal power across adjacent frequency bins, and intensity scintillations, which can help or hinder detection on characteristic timescales. In the interstellar medium, the degree of broadening and the timescale of scattering modulations are both dependent on line of sight (LOS) relative to the Galactic plane, distance traversed towards the Galactic Center (GC), and total amount of scattering material along the LOS. Though some of our LOS cut close to the Galactic plane, the GC lies below the cutoff for our GBT sample ($-20 \degr$ declination). However, our search targets the centers of nearby galaxies, and therefore a signal would have to pass through the densest portion of the host galaxy's ISM for us to detect it. Still, even in the worst-case scenario of $\sim 1$\,Hz scatter broadening for the Milky Way \citep{CordesLazio:1991}, tests show \tseti{} successfully detects signals with the correct modulation and sufficient brightness. Therefore, scatter broadening can still be neglected for our purposes. 

The greater obstacle to a narrowband microwave SETI search is intermittency. \citet{CordesLazio:1991} demonstrate that scintillation due to the ISM can modulate a signal above the detection threshold or attenuate it below. The scintillated intensities follow an exponential distribution such that the signal is attenuated most of the time. Scintillation occurs on longer timescales farther from, and for shorter paths through, the Galactic plane. Taking observations many characteristic timescales apart can lead to non-repeatability of signal detections if amplitude modulations are significant enough, but the likelihood of detection for a strongly scattered signal can be increased by taking many repeated observations of the same location, separated by periods of time longer than a characteristic scintillation timescale \citep{Cordes_1997}. Work is ongoing to better understand the impact of scintillation on narrowband radio SETI surveys and develop strategies for the detection of scintillating signals \citep{Brzycki_2023}. SETI surveys of the GC have applied the method of repeated observations \citep{Gajjar-2021}, but the observing overhead and storage requirements for this sample prevented its application for our survey. Nevertheless, we do have repeat data for a number of targets in our sample. 

Since our search deals with nearby galaxies, the intergalactic medium (IGM) and the ISMs of other galaxies also contribute to scattering in our sample. Intergalactic scattering has been comparatively less studied; studies of extragalactic phenomena such as pulsars and FRBs consider the effects of the ISM of the host galaxy as well as passage through intervening galaxies \citep{Ocker_2022}. As our sample consists of only the nearest galaxies and the beam of the GBT is sufficiently narrow, we avoid the problem of overlapping galaxies, but we note the possibility of further modulation and a shortening of the scintillation timescale as a result of scattering in the ISM of the host galaxy and the IGM between us and each galaxy.

\section{Conclusions}

We searched for technosignatures originating from 97 nearby galaxies from the \citet{isaacson:2017} sample, using four receivers covering 1.1--2.7\,GHz and 4.0--11.2\,GHz. We searched for narrow-band, Doppler drifting signals with drift rates in the range $\pm 4$\,Hz\,s$^{-1}$, with intensities above a minimum S/N threshold of 33. After algorithmic processing, correlation of signal characteristics with known RFI populations, and extensive visual inspection, we found no compelling candidate signals that were not attributable to RFI among the 1,519 events that passed our filters. 

Our search marks the one of the largest and broadest searches for radio evidence of extraterrestrial intelligence ever undertaken, surveying trillions of stars at four frequency bands. We cover the largest number of stars of any targeted radio technosignature search to date, and present the deepest search yet for high-power continuous beacons from nearby galaxies. Our fields also contain background stars and galaxies not directly targeted in this search, and accounting for their additional stars in future work could result in even more stringent constraints. Re-analysis of our sample using a modified pipeline or machine learning tools such as those developed by \citet{Ma_2023} could potentially reveal technosignatures missed by our current tools. Nevertheless, we present new techniques for the identification and investigation of the RFI environment, and document sensitivity challenges for \tseti{} to illuminate the results of past surveys.

\section{Acknowledgements}

The Breakthrough Prize Foundation funds the Breakthrough Initiatives which manages Breakthrough Listen. The Green Bank Observatory facility is supported by the National Science Foundation, and is operated by Associated Universities, Inc. under a cooperative agreement. We thank the staff at Green Bank Observatory for their support with operations. CC was funded as a participant in the Berkeley SETI Research Center Research Experience for Undergraduates Site, supported by the National Science Foundation under Grant No.~1950897. CC thanks all of the 2022 Berkeley SETI interns for their support and encouragement. S.Z.S. acknowledges that this material is based upon work supported by the National Science Foundation MPS-Ascend Postdoctoral Research Fellowship under Grant No. 2138147.

\appendix

\section{Target List}\label{app:targets}

\ref{tab:L_targets} contains a summary of the data sample analyzed in this work consisting of a count of the observations taken at each band for each target. All observations included in the sample were complete cadences, and the times provided are the start times of each cadence.

\begin{longtable}[c]{cccp{0.2\linewidth}}
\hline
\textbf{Band} & \textbf{Target} & \textbf{Observations} & \textbf{Time (MJD)} \\ 
\hline \hline
L & AND\_I     & 1 & 2019-12-15 5:02:24                                          \\
\endfirsthead
\endhead

L & AND\_II    & 2 & \multicolumn{1}{p{0.5\linewidth}}{2019-12-15 06:39:43 2021-07-10 12:20:53} \\
L & AND\_X     & 1 & 2019-12-15 3:25:38                                          \\
L & AND\_XI    & 1 & 2022-10-08 6:58:09                                          \\
L & AND\_XIV   & 1 & 2019-12-15 6:07:29                                          \\

\hhline{----}
\label{tab:L_targets}
\end{longtable}

\textbf{Note.} A machine-readable version of \ref{tab:L_targets} is published in its entirety in the electronic edition of the \textit{Astronomical Journal}. A portion is shown here for guidance regarding its form and content. The full table has the \textit{L-, S-, C-,} and \textit{X-}band data.

\clearpage

\section{Galaxy Sample Properties}\label{app:galaxy_properties}

\ref{tab:galaxy_properties} details the properties of the 97 galaxies analyzed in this work, including estimates for the number of stars captured by the telescope beam at each observing band.

\begin{longtable}[c]{p{0.1\textwidth}ccccccc}
\caption*{Properties of the Galaxy Sample Analyzed in This Work, Including Galaxy Morphology and Estimates for the Number of Stars Captured at Each Band} \\
Galaxy &
   \textbf{Distance}  &
   \textbf{K Mag}  &
  \textbf{Type} &
   \textbf{L-band Stars}  &
   \textbf{S-band Stars}  &
   \textbf{C-band Stars} &
   \textbf{X-band Stars} \\
& \textbf{(Mpc)} &&&&&
\endfirsthead
\endhead
\hline \hline
MESSIER031 & 0.77  & 0.9  & Spiral           & 3.26E+09 & 2.05E+09 & 2.17E+08 & 8.85E+07 \\
MESSIER033 & 0.85  & 4.1  & Spiral           & 6.34E+08 & 4.00E+08 & 1.38E+08 & 6.34E+07 \\
Dw1        & 2.8   & 5.1  & Spiral           & 1.58E+10 & 1.58E+10 & 1.58E+10 & 1.58E+10 \\
Maffei2    & 2.8   & 4.5  & Spiral           & 2.91E+10 & 2.91E+10 & 2.91E+10 & 5.82E+09 \\
NGC2403    & 3.18  & 6.1  & Spiral           & 7.61E+09 & 2.28E+09 & 3.04E+08 & 1.62E+08 \\
IC0342     & 3.28  & 4.5  & Spiral           & 4.12E+10 & 4.12E+10 & 4.12E+10 & 1.24E+10 \\
MESSIER081 & 3.63  & 3.8  & Spiral           & 1.05E+11 & 3.14E+10 & 2.33E+09 & 2.23E+09 \\
NGC4826    & 4.37  & 5.3  & Spiral           & 3.44E+10 & 3.44E+10 & 3.44E+10 & 3.44E+10 \\
NGC4244    & 4.49  & 7.7  & Spiral           & 3.21E+09 & 3.21E+09 & 4.59E+08 & 2.14E+08 \\
NGC4736    & 4.66  & 5.1  & Spiral           & 4.86E+10 & 4.86E+10 & 4.86E+10 & 4.86E+10 \\
NGC6503    & 5.27  & 7.3  & Spiral           & 6.85E+09 & 6.85E+09 & 6.85E+09 & 1.37E+09 \\
NGC6946    & 5.89  & 5.3  & Spiral           & 6.64E+10 & 6.64E+10 & 1.33E+10 & 7.38E+09 \\
NGC3344    & 6.85  & 7.4  & Spiral           & 1.10E+10 & 1.10E+10 & 1.10E+10 & 1.58E+09 \\
NGC0672    & 7.2   & 8.5  & Spiral           & 4.04E+09 & 4.04E+09 & 1.35E+09 & 5.77E+08 \\
NGC0628    & 7.31  & 6.8  & Spiral           & 2.34E+10 & 2.34E+10 & 2.34E+10 & 3.34E+09 \\
NGC4600    & 7.35  & 9.8  & S0               & 1.13E+09 & 1.13E+09 & 1.13E+09 & 1.13E+09 \\
MESSIER101 & 7.38  & 5.5  & Spiral           & 8.91E+10 & 8.91E+10 & 8.91E+10 & 8.91E+10 \\
NGC2787    & 7.48  & 7.2  & Spiral           & 1.64E+10 & 1.64E+10 & 1.64E+10 & 1.64E+10 \\
NGC5195    & 7.66  & 6.2  & S0               & 4.76E+10 & 4.76E+10 & 4.76E+10 & 4.76E+10 \\
NGC2683    & 7.73  & 6.3  & Spiral           & 4.39E+10 & 4.39E+10 & 4.39E+10 & 6.26E+09 \\
NGC4258    & 7.83  & 5.4  & Spiral           & 1.12E+11 & 1.12E+11 & 1.12E+11 & 1.12E+11 \\
NGC4136    & 7.9   & 9.3  & Spiral           & 2.20E+09 & 2.20E+09 & 2.20E+09 & 2.20E+09 \\
NGC7640    & 7.9   & 8.6  & Spiral           & 4.47E+09 & 4.47E+09 & 2.24E+09 & 6.39E+08 \\
NGC4618    & 7.9   & 8.6  & Spiral           & 4.47E+09 & 4.47E+09 & 4.47E+09 & 6.39E+08 \\
NGC4559    & 8.1   & 7.5  & Spiral           & 1.44E+10 & 1.44E+10 & 2.06E+09 & 6.86E+08 \\
NGC5194    & 8.4   & 5.5  & Spiral           & 1.18E+11 & 1.18E+11 & 1.18E+11 & 1.18E+11 \\
UGCA127    & 8.5   & 8.3  & Spiral           & 7.12E+09 & 7.12E+09 & 7.12E+09 & 3.56E+09 \\
Maffei1    & 3.01  & 5.5  & Elliptical       & 1.24E+10 & 1.24E+10 & 2.48E+09 & 6.52E+08 \\
NGC3379    & 11.12 & 6.2  & Elliptical       & 1.08E+11 & 1.08E+11 & 1.08E+11 & 1.08E+11 \\
NGC1400    & 24.5  & 7.8  & Elliptical       & 1.21E+11 & 1.21E+11 & 1.21E+11 & 1.21E+11 \\
LeoT       & 0.42  & 14   & Irregular        & 2.96E+04 & 2.96E+04 & 2.96E+04 & 2.96E+04 \\
NGC6822    & 0.5   & 6    & Irregular        & 1.44E+08 & 1.44E+08 & 7.57E+06 & 3.89E+06 \\
LGS 3      & 0.65  & 12.4 & Irregular        & 3.91E+05 & 3.91E+05 & 3.91E+05 & 3.91E+05 \\
IC0010     & 0.66  & 6.5  & Irregular        & 1.60E+08 & 1.60E+08 & 2.28E+07 & 8.40E+06 \\
IC1613     & 0.73  & 7.4  & Irregular        & 8.00E+07 & 2.00E+07 & 2.96E+06 & 1.63E+06 \\
Pegasus    & 0.76  & 9.8  & Irregular        & 7.69E+06 & 7.69E+06 & 7.69E+06 & 1.10E+06 \\
LeoA       & 0.81  & 10.4 & Irregular        & 4.82E+06 & 4.82E+06 & 4.82E+06 & 4.82E+06 \\
DDO210     & 0.94  & 11.2 & Irregular        & 2.97E+06 & 2.97E+06 & 2.97E+06 & 2.97E+06 \\
WLM        & 0.97  & 9    & Irregular        & 2.96E+07 & 2.96E+07 & 9.86E+06 & 5.91E+06 \\
Sag dIr    & 1.04  & 12.1 & Irregular        & 1.49E+06 & 1.49E+06 & 1.49E+06 & 2.98E+05 \\
SexA       & 1.32  & 10.1 & Irregular        & 1.91E+07 & 1.91E+07 & 1.91E+07 & 1.91E+06 \\
SexB       & 1.36  & 9.5  & Irregular        & 3.75E+07 & 3.75E+07 & 3.75E+07 & 5.35E+06 \\
UGC04879   & 1.36  & 11.5 & Irregular        & 4.94E+06 & 4.94E+06 & 4.94E+06 & 4.94E+06 \\
HIZSS003   & 1.67  & 11.3 & Irregular        & 9.51E+06 & 9.51E+06 & 9.51E+06 & 9.51E+06 \\
UMin       & 0.06  & 7.6  & Dwarf Spheroidal & 3.83E+04 & 1.28E+04 & 5.36E+03 & 5.47E+03 \\
Draco      & 0.08  & 6.6  & Dwarf Spheroidal & 1.54E+05 & 4.79E+04 & 2.96E+04 & 2.96E+04 \\
SexDSph    & 0.09  & 7    & Dwarf Spheroidal & 1.20E+06 & 1.20E+06 & 1.20E+06 & 1.20E+06 \\
Hercules   & 0.15  & 10.6 & Dwarf Spheroidal & 9.63E+04 & 9.63E+04 & 9.63E+04 & 9.63E+04 \\
LeoII      & 0.21  & 8.4  & Dwarf Spheroidal & 1.87E+06 & 1.87E+06 & 1.87E+06 & 1.87E+06 \\
CVnI       & 0.22  & 9.7  & Dwarf Spheroidal & 5.56E+05 & 5.56E+05 & 5.56E+05 & 5.56E+05 \\
And\_XVI   & 0.52  & 11.3 & Dwarf Spheroidal & 7.30E+05 & 7.30E+05 & 7.30E+05 & 7.30E+05 \\
And\_XXIV  & 0.6   & 12.7 & Dwarf Spheroidal & 2.42E+05 & 2.42E+05 & 2.42E+05 & 2.42E+05 \\
NGC0185    & 0.61  & 6.5  & Dwarf Spheroidal & 1.34E+08 & 6.71E+07 & 7.06E+06 & 2.74E+06 \\
Bol520     & 0.63  & 13.3 & Dwarf Spheroidal & 1.47E+05 & 1.47E+05 & 1.47E+05 & 1.47E+05 \\
And\_X     & 0.63  & 12.1 & Dwarf Spheroidal & 4.95E+05 & 4.95E+05 & 4.95E+05 & 4.95E+05 \\
And\_II    & 0.65  & 10.7 & Dwarf Spheroidal & 2.19E+06 & 2.19E+06 & 2.19E+06 & 2.19E+06 \\
And\_XIV   & 0.73  & 12.5 & Dwarf Spheroidal & 4.56E+05 & 4.56E+05 & 4.56E+05 & 4.56E+05 \\
And\_XI    & 0.73  & 14   & Dwarf Spheroidal & 9.98E+04 & 9.98E+04 & 9.98E+04 & 9.98E+04 \\
And\_I     & 0.73  & 9.5  & Dwarf Spheroidal & 9.53E+06 & 9.53E+06 & 9.53E+06 & 9.53E+06 \\
And XXIII  & 0.73  & 10.7 & Dwarf Spheroidal & 2.83E+06 & 2.83E+06 & 2.83E+06 & 2.83E+06 \\
NGC4489    & 15.2  & 9.4  & Elliptical       & 8.40E+09 & 8.40E+09 & 8.40E+09 & 8.40E+09 \\
NGC4486B   & 15.4  & 10.1 & Elliptical       & 4.25E+09 & 4.25E+09 & 4.25E+09 & 4.25E+09 \\
MESSIER59  & 15.5  & 6.7  & Elliptical       & 1.35E+11 & 1.35E+11 & 1.35E+11 & 1.35E+11 \\
MESSIER49  & 15.81 & 5.4  & Elliptical       & 5.27E+11 & 5.27E+11 & 5.27E+11 & 7.53E+10 \\
NGC4478    & 16    & 8.4  & Elliptical       & 2.59E+10 & 2.59E+10 & 2.59E+10 & 2.59E+10 \\
MESSIER86  & 16.08 & 6.1  & Elliptical       & 2.69E+11 & 2.69E+11 & 2.99E+10 & 1.17E+10 \\
NGC4473    & 16.18 & 7.2  & Elliptical       & 8.95E+10 & 8.95E+10 & 8.95E+10 & 2.98E+10 \\
NGC4660    & 16.4  & 8.2  & Elliptical       & 3.35E+10 & 3.35E+10 & 3.35E+10 & 3.35E+10 \\
MESSIER60  & 16.55 & 5.7  & Elliptical       & 4.30E+11 & 4.30E+11 & 4.30E+11 & 4.30E+11 \\
MESSIER87  & 16.6  & 5.8  & Elliptical       & 3.91E+11 & 3.91E+11 & 3.91E+11 & 3.91E+11 \\
MESSIER84  & 16.7  & 6.2  & Elliptical       & 2.64E+11 & 2.64E+11 & 8.81E+10 & 3.30E+10 \\
NGC4564    & 16.9  & 7.9  & Elliptical       & 4.85E+10 & 4.85E+10 & 4.85E+10 & 4.85E+10 \\
NGC4551    & 16.9  & 8.9  & Elliptical       & 1.76E+10 & 1.76E+10 & 1.76E+10 & 1.76E+10 \\
NGC4387    & 17    & 9.2  & Elliptical       & 1.32E+10 & 1.32E+10 & 1.32E+10 & 1.32E+10 \\
NGC4239    & 17.4  & 10.2 & Elliptical       & 5.03E+09 & 5.03E+09 & 5.03E+09 & 5.03E+09 \\
NGC4458    & 17.6  & 9.3  & Elliptical       & 1.28E+10 & 1.28E+10 & 1.28E+10 & 1.28E+10 \\
NGC584     & 19.09 & 7.3  & Elliptical       & 1.16E+11 & 1.16E+11 & 1.16E+11 & 1.16E+11 \\
NGC1052    & 19.72 & 7.5  & Elliptical       & 1.02E+11 & 1.02E+11 & 1.02E+11 & 1.02E+11 \\
NGC596     & 21.08 & 8    & Elliptical       & 7.12E+10 & 7.12E+10 & 7.12E+10 & 7.12E+10 \\
NGC7454    & 21.14 & 8.9  & Elliptical       & 2.88E+10 & 2.88E+10 & 2.88E+10 & 2.88E+10 \\
NGC4365    & 21.4  & 6.6  & Elliptical       & 3.04E+11 & 3.04E+11 & 7.60E+10 & 2.34E+10 \\
NGC4434    & 22.1  & 9.2  & Elliptical       & 2.34E+10 & 2.34E+10 & 2.34E+10 & 2.34E+10 \\
NGC720     & 23.18 & 7.3  & Elliptical       & 1.78E+11 & 1.78E+11 & 1.78E+11 & 1.78E+11 \\
NGC1407    & 23.26 & 6.7  & Elliptical       & 3.30E+11 & 3.30E+11 & 3.30E+11 & 3.30E+10 \\
NGC1172    & 23.7  & 9.2  & Elliptical       & 2.73E+10 & 2.73E+10 & 2.73E+10 & 2.73E+10 \\
NGC821     & 23.92 & 7.9  & Elliptical       & 1.04E+11 & 1.04E+11 & 1.04E+11 & 1.04E+11 \\
NGC5638    & 24    & 8.3  & Elliptical       & 6.99E+10 & 6.99E+10 & 6.99E+10 & 6.99E+10 \\
NGC5322    & 24.1  & 9.9  & Elliptical       & 1.39E+10 & 1.39E+10 & 1.39E+10 & 1.39E+10 \\
NGC4318    & 24.16 & 10.3 & Elliptical       & 9.35E+09 & 9.35E+09 & 9.35E+09 & 9.35E+09 \\
NGC636     & 24.9  & 8.4  & Elliptical       & 6.85E+10 & 6.85E+10 & 6.85E+10 & 6.85E+10 \\
NGC5846    & 26.9  & 6.9  & Elliptical       & 3.71E+11 & 3.71E+11 & 3.71E+11 & 2.97E+11 \\
NGC3193    & 27.5  & 8    & Elliptical       & 1.28E+11 & 1.28E+11 & 1.28E+11 & 1.28E+11 \\
NGC5831    & 27.5  & 8.4  & Elliptical       & 8.52E+10 & 8.52E+10 & 8.52E+10 & 8.52E+10 \\
NGC5845    & 27.6  & 9.1  & Elliptical       & 4.23E+10 & 4.23E+10 & 4.23E+10 & 4.23E+10 \\
NGC4168    & 28.1  & 8.4  & Elliptical       & 8.94E+10 & 8.94E+10 & 8.94E+10 & 8.94E+10 \\
NGC3226    & 29.1  & 8.6  & Elliptical       & 7.88E+10 & 7.88E+10 & 7.88E+10 & 7.88E+10 \\
NGC5813    & 29.2  & 7.6  & Elliptical       & 2.19E+11 & 2.19E+11 & 2.19E+11 & 2.19E+11 \\
\hline
\label{tab:galaxy_properties}\\
\end{longtable}

\textbf{Note.} Modified from \citet{isaacson:2017}; further properties are recorded there. Stellar number estimates were calculated using the \textit{K-}band magnitudes listed, via the methods of \citet{Ma_2014} (see Equations 1 and 2). Stars were assumed to have a mass of one $M_\sun$. This table is available in machine-readable form.

\clearpage

\section{DBSCAN Outliers}\label{app:outliers}

\begin{figure*}[!hpt] 
\centering

\subfigure[]{
  \includegraphics[width=0.45\textwidth]{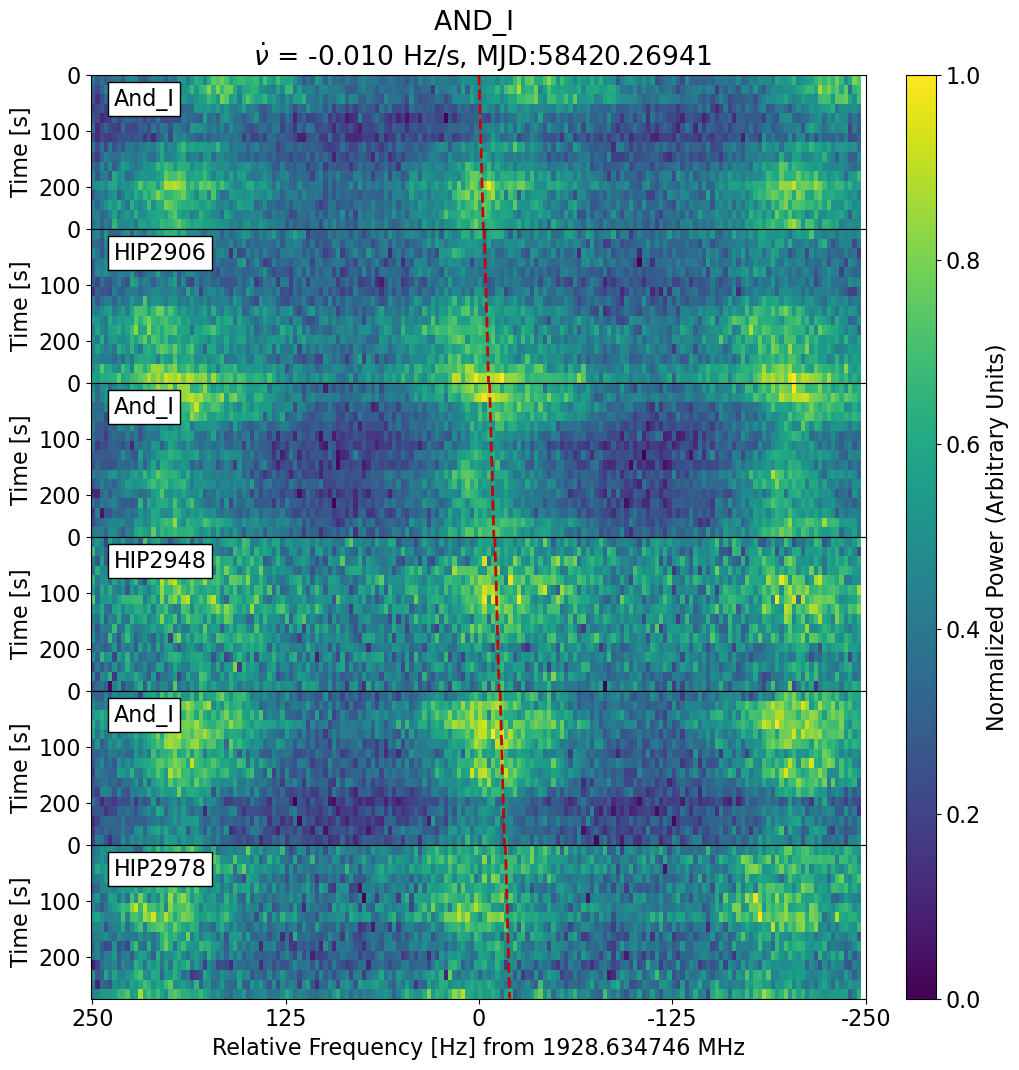} 
  \label{fig:int_event2}
}
\hfill
\subfigure[]{
  \includegraphics[width=0.45\textwidth]{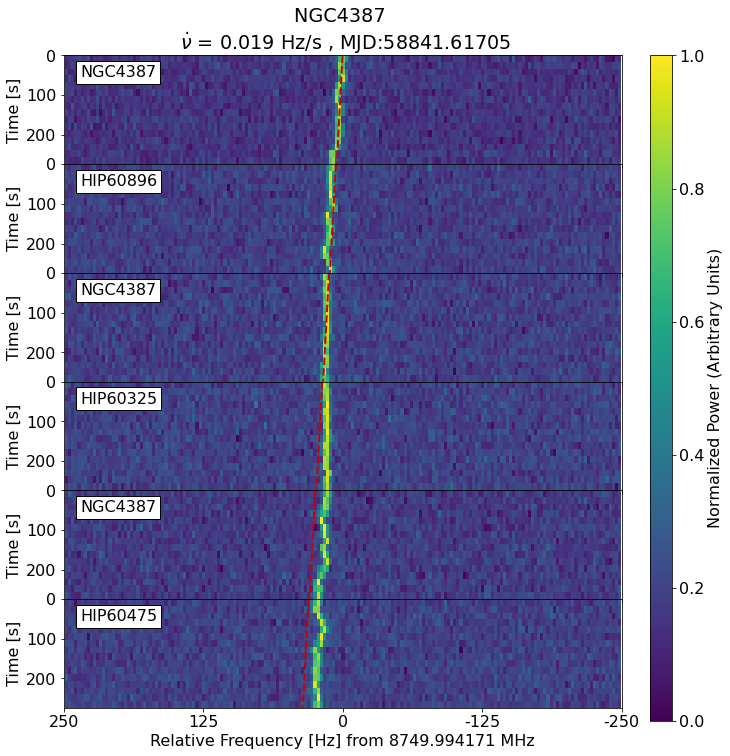}
  \label{fig:int_event3}
}

\subfigure[]{
  \includegraphics[width=0.45\textwidth]{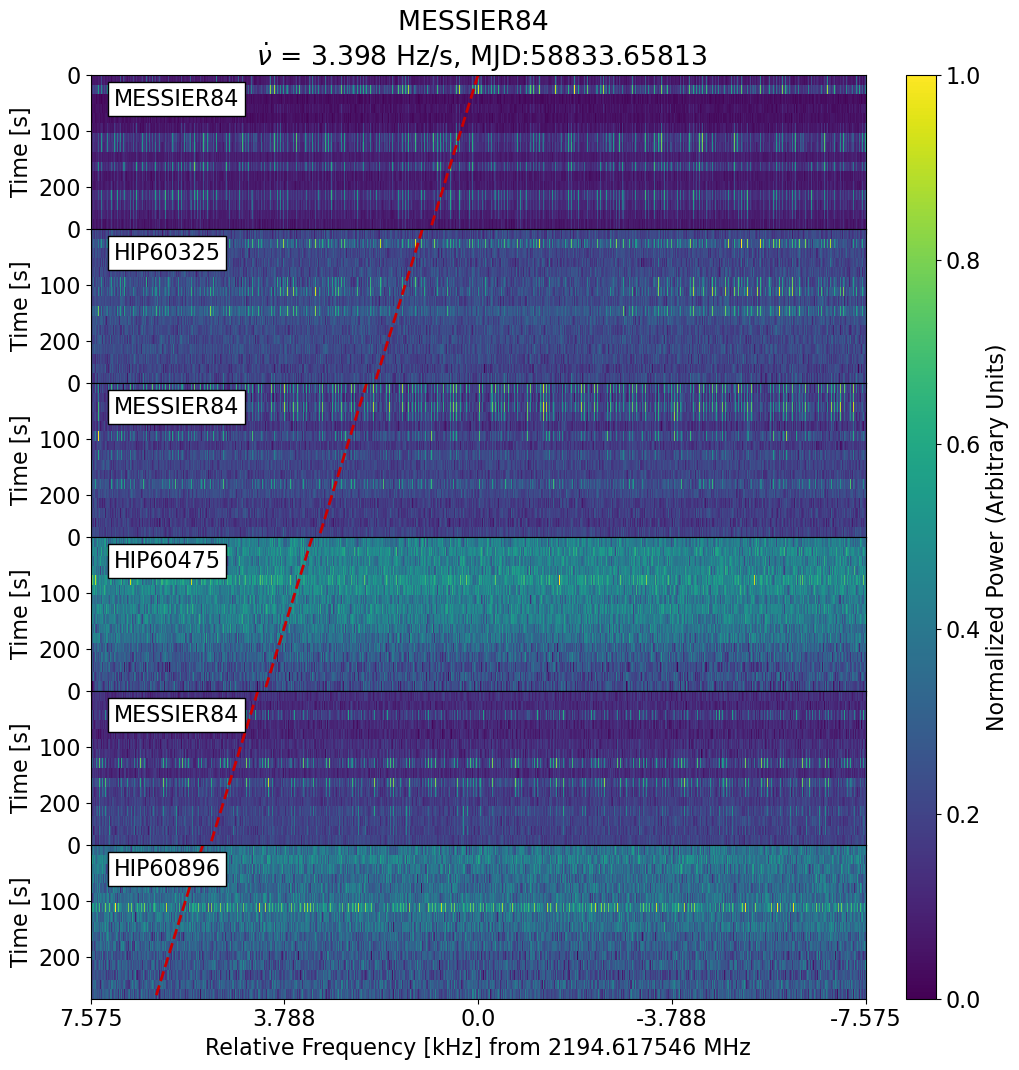}
  \label{fig:int_event1}
}
\hfill
\subfigure[]{
  \includegraphics[width=0.45\textwidth]{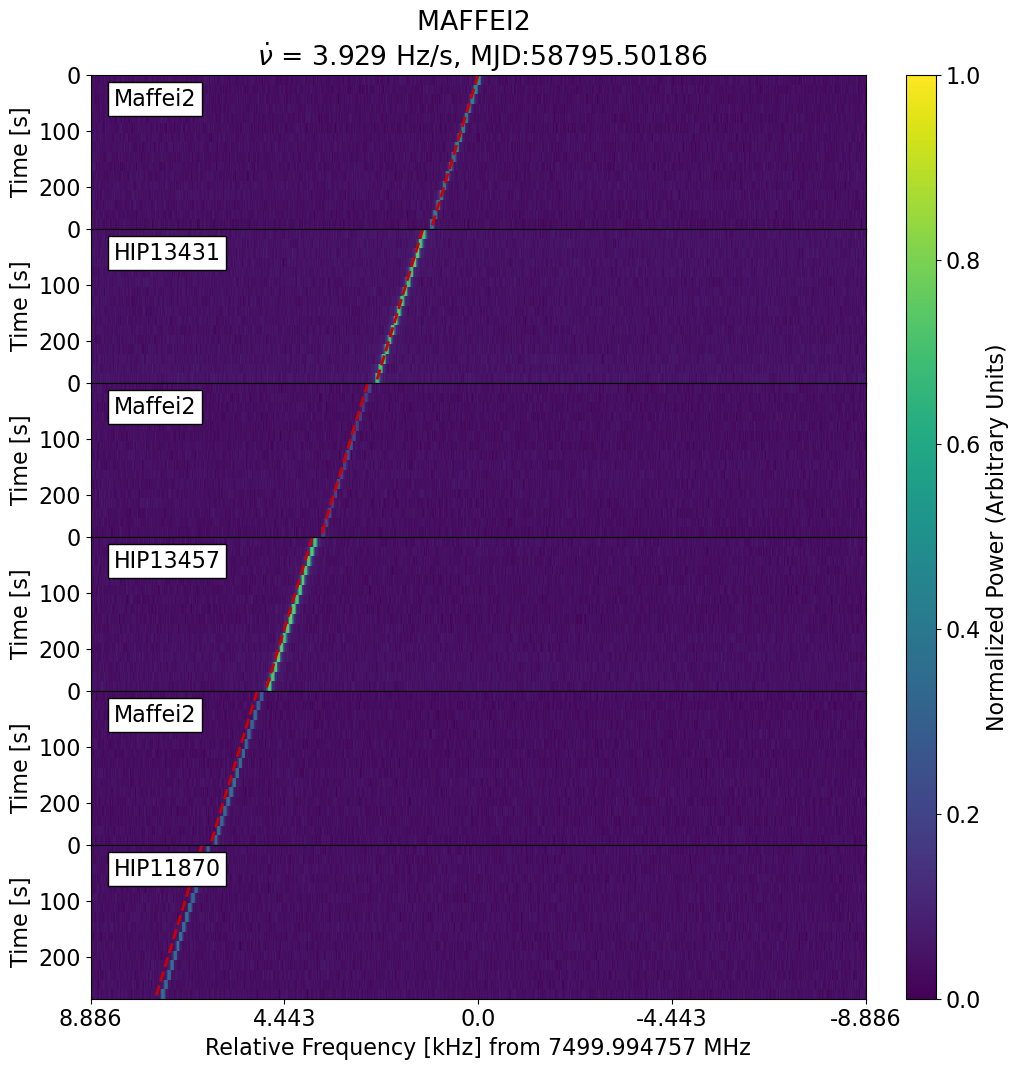}
  \label{fig:int_event5}
}

\caption{Dynamic spectra (waterfall plots) of four representative events from the 634 event sample. Each plot is a vertical stack of the 6 scans making up an ABACAD cadence. The vertical axis shows the time since the start of each scan in the cadence, and the horizontal axis shows the frequency offset from the event's starting frequency.}
\label{fig:int_events}
\end{figure*}

Figure \ref{fig:int_events} includes four dynamic spectra of events flagged as outliers by DBSCAN clustering.

\clearpage

\section{DBSCAN Clusters in Context}\label{app:rfi_context}

\begin{figure*}[!hpt]
\centering

\subfigure[]{
  \includegraphics[width=0.45\columnwidth]{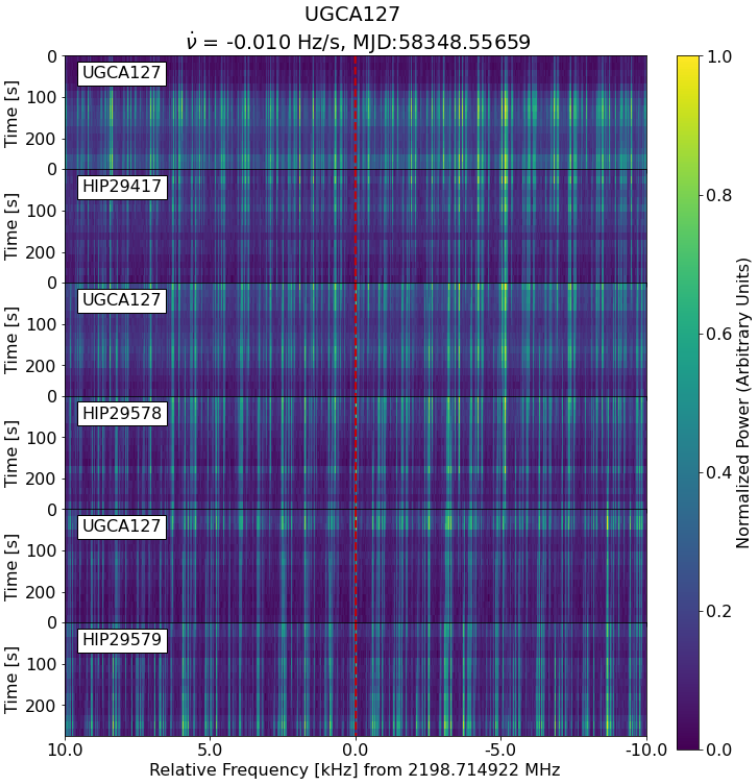} 
  \label{fig:cluster_1_zoom}
}
\hfill
\subfigure[]{
  \includegraphics[width=0.45\columnwidth]{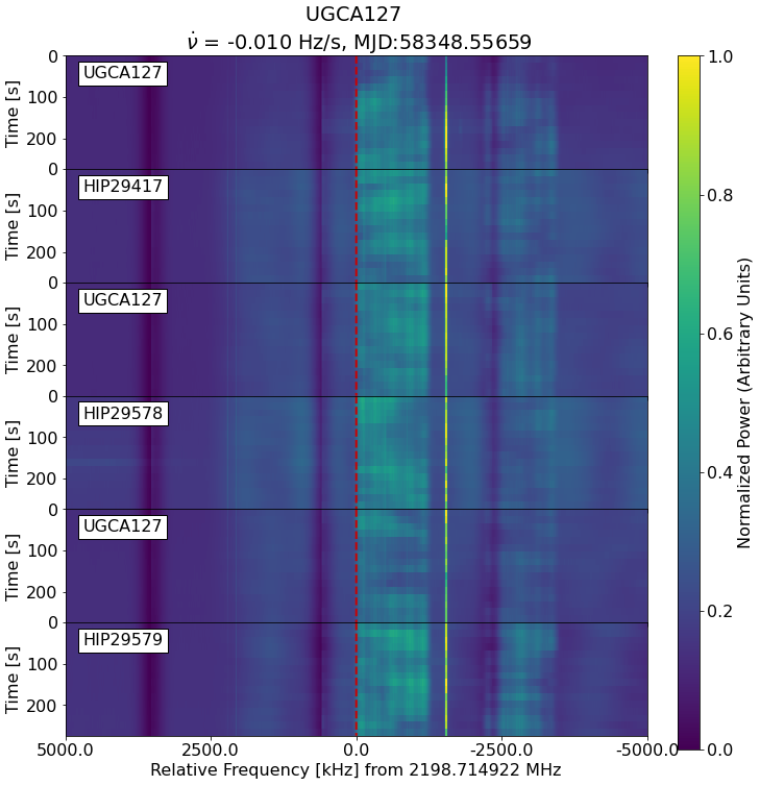}
  \label{fig:cluster_1}
}

\subfigure[]{
  \includegraphics[width=0.45\columnwidth]{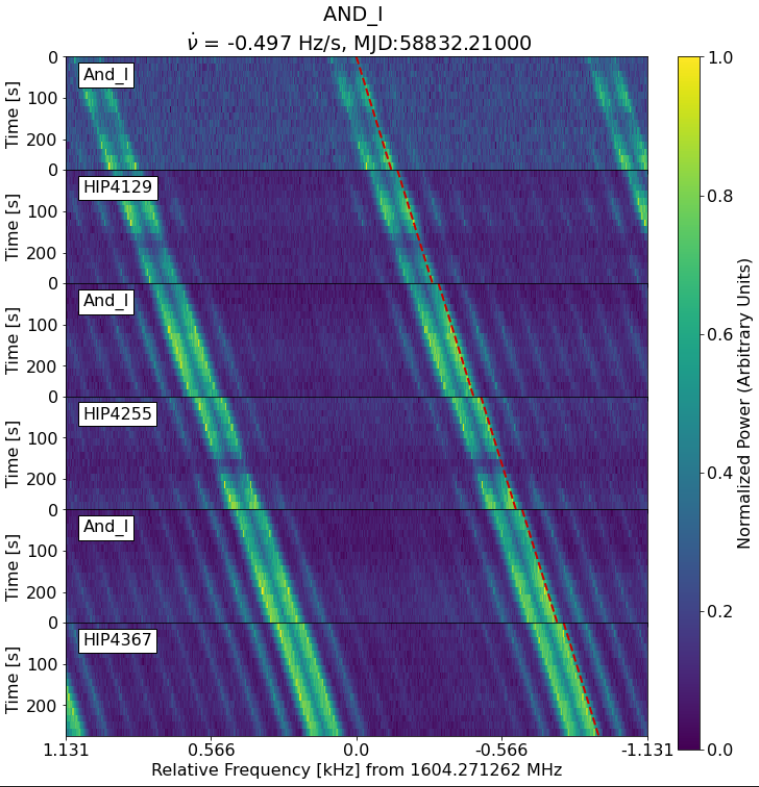}
  \label{fig:cluster_13_zoom}
}
\hfill
\subfigure[]{
  \includegraphics[width=0.45\columnwidth]{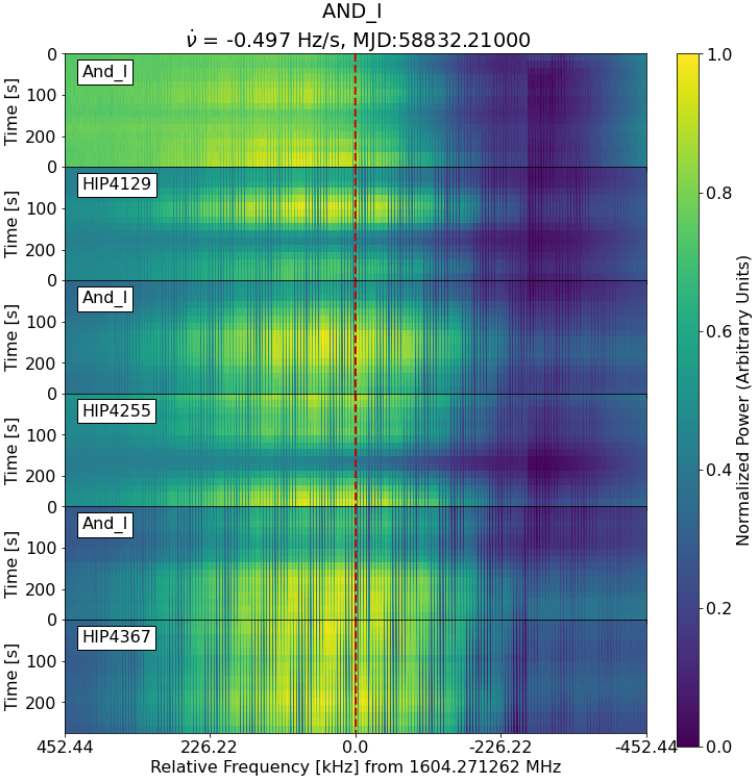}
  \label{fig:cluster_13}
}

\caption{Dynamic spectra (waterfall plots) of events from clusters 1 and 13 (left column), zoomed out to show RFI context (right column). Both events are accompanied by high-density ensembles of parallel signals across wide bandwidths. Cluster 1 spans several different sources of broadband RFI with close to zero drift, including the ensemble seen in the zoom-out of cluster 13, but with densely spaced broadband phenomena, lines brighter than the local environment can be selected for many drift rates. The events depicted in (a) and (b) are in a frequency range used for satellite communications, and the events depicted in (c) and (d) match the Global Navigation Satellite System band in frequency range.}
\label{fig:clusters_1_13}
\end{figure*}

\begin{figure*}[!hpt]
\centering

\subfigure[]{
  \includegraphics[width=0.45\columnwidth]{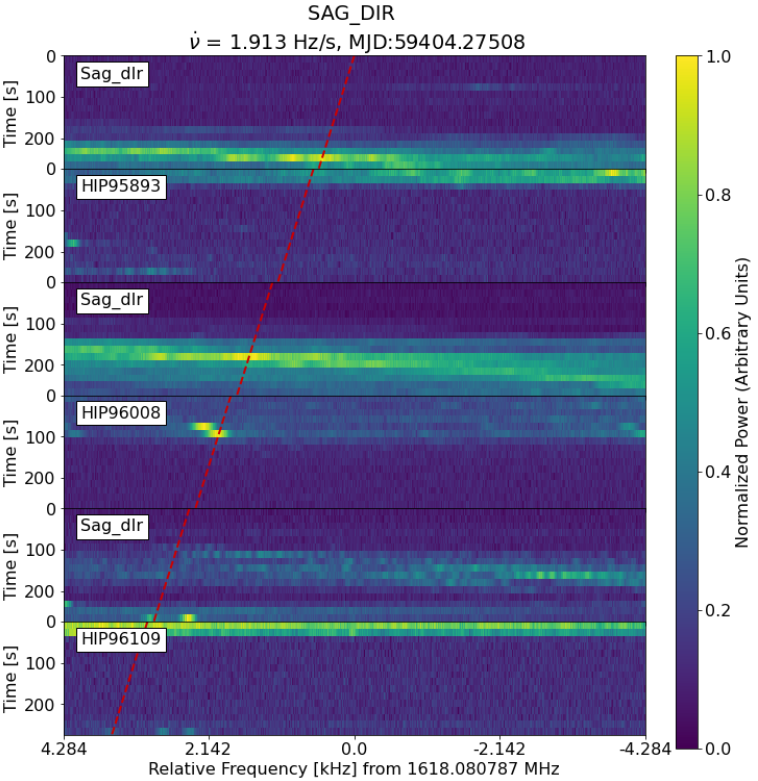} 
  \label{fig:cluster_14_zoomed_in}
}
\hfill
\subfigure[]{
  \includegraphics[width=0.45\columnwidth]{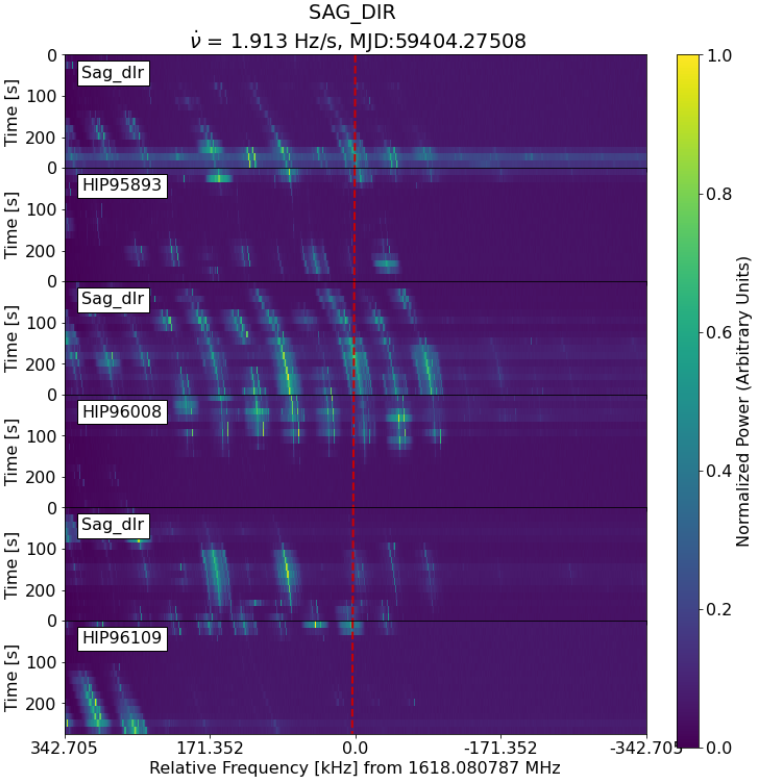}
  \label{fig:cluster_14_zoomed_out}
}

\subfigure[]{
  \includegraphics[width=0.45\columnwidth]{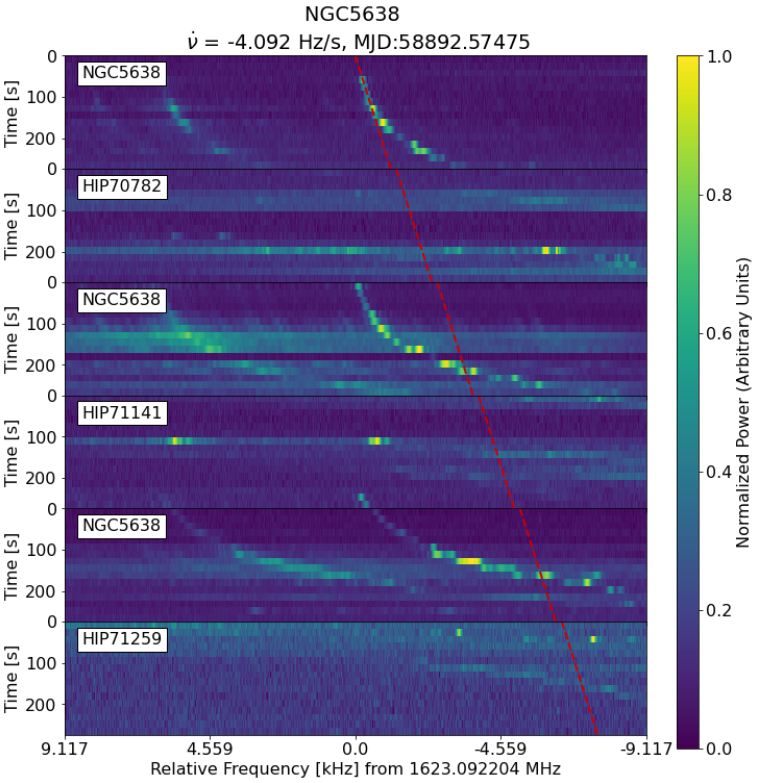}
  \label{fig:cluster_15_zoomed_in}
}
\hfill
\subfigure[]{
  \includegraphics[width=0.45\columnwidth]{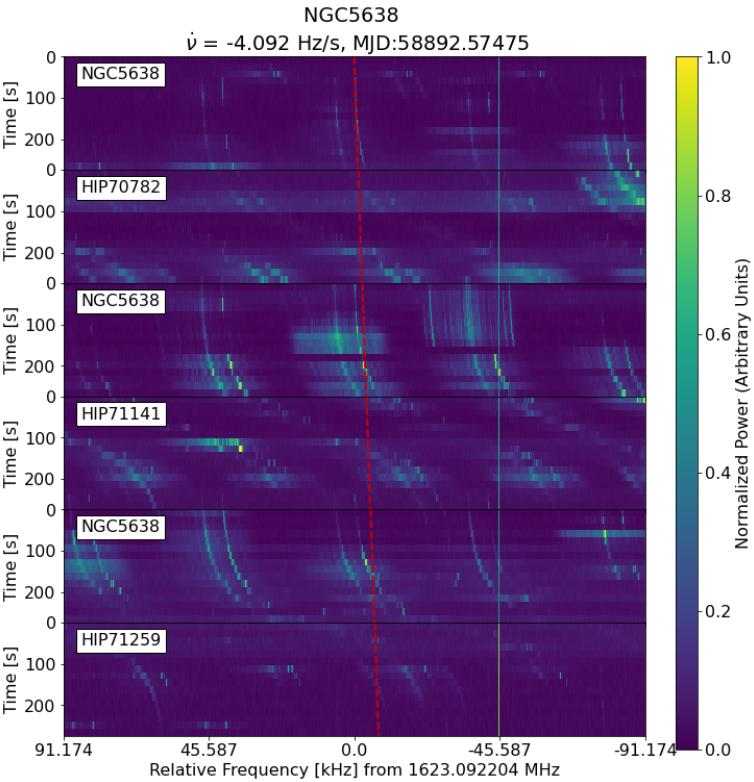}
  \label{fig:cluster_15_zoomed_out}
}

\caption{Dynamic spectra (waterfall plots) of events from clusters 14 and 15, zoomed out to show RFI context. The two events are close in frequency but far in drift rate, and part of collections of periodically drifting RFI. Both frequencies lie within allocations for aeronautical radionavigation and mobile satellites for \textit{L} band.}
\label{fig:cluster_14_15}
\end{figure*}

Figures \ref{fig:clusters_1_13} and \ref{fig:cluster_14_15} include dynamic spectra of events in the largest DBSCAN clusters and their RFI contexts. All large clusters map clearly to frequency ranges associated with anthropogenic transmissions.

\clearpage

\nocite{Phoenix_all:2004, Tremblay:2020, Horowitz:1993, Harp:2016, Siemion:2013}


\bibliographystyle{aasjournal}
\bibliography{main.bib}

\end{document}